\documentclass[preprint]{aastex}        % single-spaced version  USE THIS ONE FOR FINAL

\usepackage{lscape}
\usepackage{epsfig,here,rotate}   
 
\newcommand{\vinfty}{$v_{\infty}\ $}
\newcommand{\vimp}{$v_{imp}\ $}
\newcommand{\degrees}{$^\circ\ $}                 % degrees (in text)
\newcommand{\eg}{{\it e.g.,}$\;$}

\newcommand{\etal}{{\it et~al.}}
\newcommand{\beq}{\begin{equation}}
\newcommand{\eeq}{\end{equation}}
 
\shorttitle{}
\singlespace

\begin{document}
 
\title{Current Bombardment of the Earth-Moon System: 
\\Emphasis on Cratering Asymmetries}
 
%}
\author{J.~Gallant, B.~Gladman, and M.~\'{C}uk$^1$} 
\affil{
$^1$Department of Physics and Astronomy, University of British Columbia, \\
6224 Agricultural Road, Vancouver, BC V6T 1Z1 Canada}
%\author{Matija ~\'{C}uk} %\affil{
%Department of Physics and Astronomy, University of British Columbia, \\
%6224 Agricultural Road, Vancouver, BC V6T 1Z1 Canada 
%}
\maketitle

%-----------------------------------------------------

%\newpage
 
\begin{center} 
{\bf Abstract}
\end{center}

We calculate the current spatial distribution of projectile delivery 
to the Earth and Moon using numerical orbital dynamics simulations of
candidate impactors drawn from a debiased Near-Earth-Object (NEO)
model.  Surprisingly, we find that the average lunar impact velocity
is 20 km/s, which has ramifications in converting observed crater densities to
impactor size distributions.  We determine that current crater
production on the leading hemisphere of the Moon is $1.29 \pm 0.01$
that of the trailing when considering the ratio of craters within
30\degrees of the apex to those within 30\degrees of the antapex and
that there is virtually no nearside-farside asymmetry.  As expected,
the degree of leading-trailing asymmetry increases when the Moon's
orbital distance is decreased.  We examine the latitude distribution
of impactor sites and find that for both the Earth and Moon there is a
small deficiency of time-averaged impact rates at the poles.  The
ratio between deliveries within 30\degrees of the pole to that of a
30\degrees band centered on the equator is nearly unity for Earth ($<
1\%$)($0.992 \pm 0.001$) but detectably non-uniform for the Moon
($\sim 10\%$) ($0.912 \pm 0.004$).  The terrestrial arrival results
are examined to determine the degree of AM/PM asymmetry to compare
with meteorite fall times (of which there seems to be a PM excess).
Our results show that the impact flux of objects derived from the NEOs
in the AM hours is $\sim$ 2 times that of the PM hemisphere, further
supporting the assertion that meteorite-dropping objects are recent
ejections from the main asteroid belt rather than young fragments of
NEOs.  

Submitted to {\sl Icarus}: July 28 2006 \\
Send correspondance to B.G.: gladman@phas.ubc.ca

%-----------------------------------------------------
\newpage
 
\section{Introduction} 

The Clementine mission to survey the lunar surface, with its high
resolution imagery, has re-opened the dormant field of crater counting
on the Moon \citep{pete94,moore96,mcewen97}.  \citet{morota03}
have used these high quality images to count young rayed craters
and concluded that a factor of 1.5 enhancement exists in the current
cratering record when comparing fitted crater densities precisely at
the apex to those directly at the antapex. An excess of
leading-hemisphere craters is expected on synchronously-rotating
satellites in general; \citet{hor84} give expressions for the expected
asymmetry caused by the leading hemisphere of the satellite ``sweeping
up'' more objects.  Based on such expressions and using their previous
crater counts, \citet{morota05} concluded that the measured leading
enhancement implies an average impactor encounter speed with the
Earth-Moon system of 12-16 km/s, consistent (though on the low end)
with previous estimates \citep{shoe83,chyba94}. 

Other synchronously-rotating moons have been studied in great
detail using the high-resolution images from Voyager.  By examining
previously-obtained crater counts of the Galilean satellites,
\citet{zetal01} reconfirm the prediction of a leading hemisphere
excess (which is quite large in the Jovian system due to high satellite
orbital speeds) and are able to derive an analytical expression to
describe the asymmetry that is consistent with results of
\citet{shoe82} and \citet{hor84}.   

In addition to this leading/trailing asymmetry, some authors believe
that the Earth could act as a gravitational lens, focusing incoming
projectiles onto the nearside of the Moon, causing an asymmetry
between the near and far sides \citep{turk62,wiesel71}.  This concept
has been numerically investigated by \citet{wood73,fev05} and deemed
plausible, though an analytic calculation by \citet{band73} showed the
amplitude of the asymmetry depends on the Earth-Moon distance and the
velocity of the incoming projectiles.  

%\citet{wiesel71} introduced the idea that   This was consistent with
%an earlier two-dimensional calculation by \cite{turk62}, who found a
%factor of two enhancement in the nearside flux.  \citet{wood73} used a
%numerical calculation, to establish the potential existence of a
%nearside enhancement.  In contrast,  \citet{band73} showed
%analytically that in some situations the Earth could act not as a
%lens, but rather as a shield by blocking incoming projectiles.  Which
%of the two effects dominates depends on the Earth-Moon distance and
%the velocity of the incoming projectiles. \cite{fev05} reinvestigated
%this matter using numerical simulations and concluded in a preliminary
%analysis that the averaged number of near side impacts are enhanced by
%a factor of four when compared to the far side.  

Our goal is to determine if the present-day flux of Near-Earth Objects
(NEOs) would produce a spatial distribution of young craters on the
lunar landscape that agrees with these studies, in terms of both
leading/trailing and nearside/farside asymmetries.  Our approach
involves many N-body simulations of projectiles drawn from the
debiased NEO model of \citet{bottke02}, which are then injected into
the Earth-Moon system.   

Since a realistic impactor distribution is not expected to be isotropic,
one might find some latitudinal variation in the spatial distribution
of deliveries on both the Earth and Moon.  This concept has been studied in
the past \citep{xmas64}, but more recently by \citet{fev06} who
concluded 60\% and 30\% latitudinal variations for the Moon and Earth
respectively when comparing the impact density at the poles to that of
the equator.  We seek to compare our direct N-body simulations to
their semi-analytic approach as they also used the \citet{bottke02}
NEO model.      

Asymmetries may also be present for the Earth and would be most
reliably observed via fireball sightings and meteorite recoveries.
Based on observations, many researchers concluded that more daytime
fireballs occur in the afternoon hours, causing a morning/afternoon
(AM/PM) asymmetry \citep{wether68,xmas82}.  Several studies
\citep{xmas82,wether85,morglad98} have been done to understand this 
PM excess.  \citet{xmas82} produced a PM fireball excess, but drew from a
highly-selected group of orbital parameters for the meteoroid source
population.  These orbits may not represent the true distribution of objects in
near-Earth space. Thus, since we will obtain many simulated Earth
impacts in our lunar study, an additional benefit is to reinvestigate
the question of the AM/PM asymmetry.  The debiased NEO model we use as our source
population should give a better approximation of the objects
in near-Earth space than the parameters used by \citet{xmas82}.
Annual effects are also of interest as at different times of the year
certain locations on Earth receive an enhancement in the
impactor flux due to the location of the Earth's spin axis relative
to the incoming flux direction \citep{xmas82,renkno89}.

In Section~\ref{sec-theory}, we give a brief overview on the theory of
various asymmetries in the Earth-Moon system.  Section~\ref{sec-MM}
describes the model for our simulations as well as the methods of
implementation.  Our findings begin in Section~\ref{sec-ED}, where we
first examine deliveries of projectiles to Earth in terms of 
the latitude distribution and then the AM/PM asymmetry.  Lunar results
follow in Section~\ref{sec-LB} where we examine the latitude 
distribution and the leading/trailing and nearside/farside hemispheric
asymmetries.  Section~\ref{sec-con} summarizes our key findings and
discusses some possible implications.

\section{Asymmetry Theory}
\label{sec-theory}

Why search for asymmetries?  For the Earth in
terms of a morning/afternoon asymmetry, one can glean some information
about the origin of the impacting population as certain types of
orbits will preferentially strike during specific local times.  More
importantly though, when looking at the the Moon, any spatial variation
in the observed crater density will affect how those craters are used to
interpret the complex cratering history of the Moon. 

\subsection{Meteorite fall statistics on Earth}
The details concerning an afternoon (PM) excess of meteorite falls on the
Earth has been the subject of much debate, creating a large body of
literature on the subject (\eg
\citet{xmas64,wether68,xmas82,wether85,morglad98} and references
therein).  This concept was instigated by the fact that chondrites
seem to be biased to fall during PM hours.  \citet{wether68}
quantified the effect as the ratio of daytime PM falls to the total
number of daytime falls and arrived at a ratio $\sim 0.68$, where the
convention is to count only falls between 6AM and 6PM since human
observers are less numerous between midnight and 6AM.  However, even
this daytime value may be socially biased as there are more potential
observers from noon-6PM than in the 6AM-noon interval.  A hypothesis
to explain a PM excess is that prograde, low-inclination meteoroids
with semi-major axes $> 1$~AU and pericentres just inside 1~AU
preferentially coat Earth's trailing hemisphere.  Several dynamical
and statistical studies \citep{xmas82,wether85,morglad98} support this
afternoon enhancement scenario.  In our work, we will use the debiased
NEO model of \citet{bottke02} to compute the distribution of fall
times from an NEO source population.  We can use this to learn about
the true orbital distribution of meteorite dropping objects.    

\subsection{Lunar leading hemisphere enhancement}
\label{sec:llhe}
A leading hemisphere enhancement originates from the satellite's
motion about its host planet.  As it orbits, the leading hemisphere
tends to encounter more projectiles, thus enhancing the crater
production on that side.  The faster the satellite orbits, the more
difficult it becomes for objects to encounter the trailing hemisphere
and the higher the leading-side impact speeds become. In addition to this
effect, the size-frequency distribution of craters will be skewed
towards larger diameter craters on the leading hemisphere.  A crater,
of size $D_c$, at the apex will have been produced from a
smaller-sized impactor (on average) than one which makes the same
sized crater at the antapex because impact speeds $v_{imp}$ are
generally higher on the leading hemisphere and the crater diameter
scales roughly as $D_c \propto v_{imp}^{2/3}$ (\eg \citet{boris01}).   

%BOOGERS

Analytic investigations into leading/trailing asymmetries have ranged
from the general \citep{hor84} to the very specific
\citep{shoe82,zetal98}, but all analytic treatments are forced make
the assumption that the impacting population has an
isotropic distribution in the rest frame of the planet.  In some
cases, the gravity of the satellite is ignored, and in the above
treatments, impact probabilities (depending on the encounter direction
and speed) were not included in the derivation of cratering rates.   

For fixed impactor encounter speed and an isotropic source
distribution, the areal crater density $\Gamma$ would follow the
functional form  
\begin{equation}
\Gamma(\beta) = \bar{\Gamma}\left(1 + \alpha\cos\beta \right)^g,
\label{eq:ass}
\end{equation}
where $\beta$ is the angle from the apex of the moon's motion and
$\bar{\Gamma}$ is the value of the crater density at $\beta =
90$\degrees \citep{zetal01,morota06}. 
The ``amplitude'' $\alpha$ of the asymmetry is related to the orbital
velocity $v_{orb}$ of the Moon and the velocity of the projectiles at
infinity \vinfty by 
\begin{equation}
\alpha = \frac{v_{orb}}{\sqrt{2v_{orb}^2 + v_{\infty}^2}}\;.
\label{eq:alf}
\end{equation}
Note that for large encounter speeds, $\alpha \rightarrow
v_{orb}/v_{\infty}$ and that $\alpha \rightarrow 1/\sqrt{2}$ as
$v_{\infty} \rightarrow 0$.

The exponent $g$ (which also effects the asymmetry) is expressed as
\begin{equation}
g= 2.0 + 1.4\;b,
\label{eq:gee}
\end{equation}
where $b$ is the slope of the cumulative mass distribution for the
impactors
\beq
N(>m) \propto m^{-b}\;.
\eeq
Based on current observations \citep{bottke02}, $b = 0.58 \pm 0.03$, making $g
= 2.81 \pm 0.05$. 

\begin{figure}[H]
\vspace{0.5cm}
%\begin{center}
\hspace{2.5cm}\epsfig{file=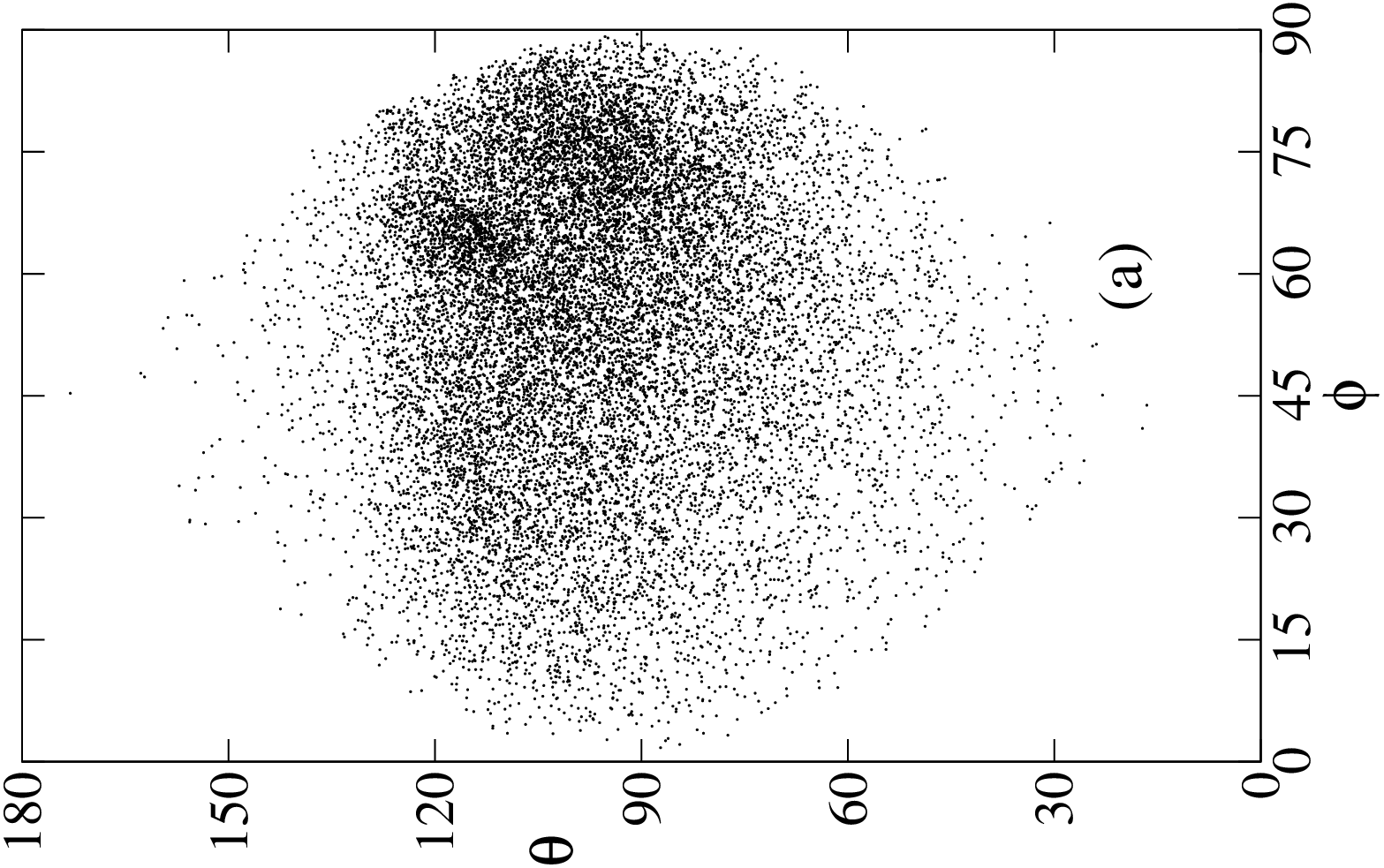,height=7.0cm,angle=-90}
\hspace{-1.5cm}\epsfig{file=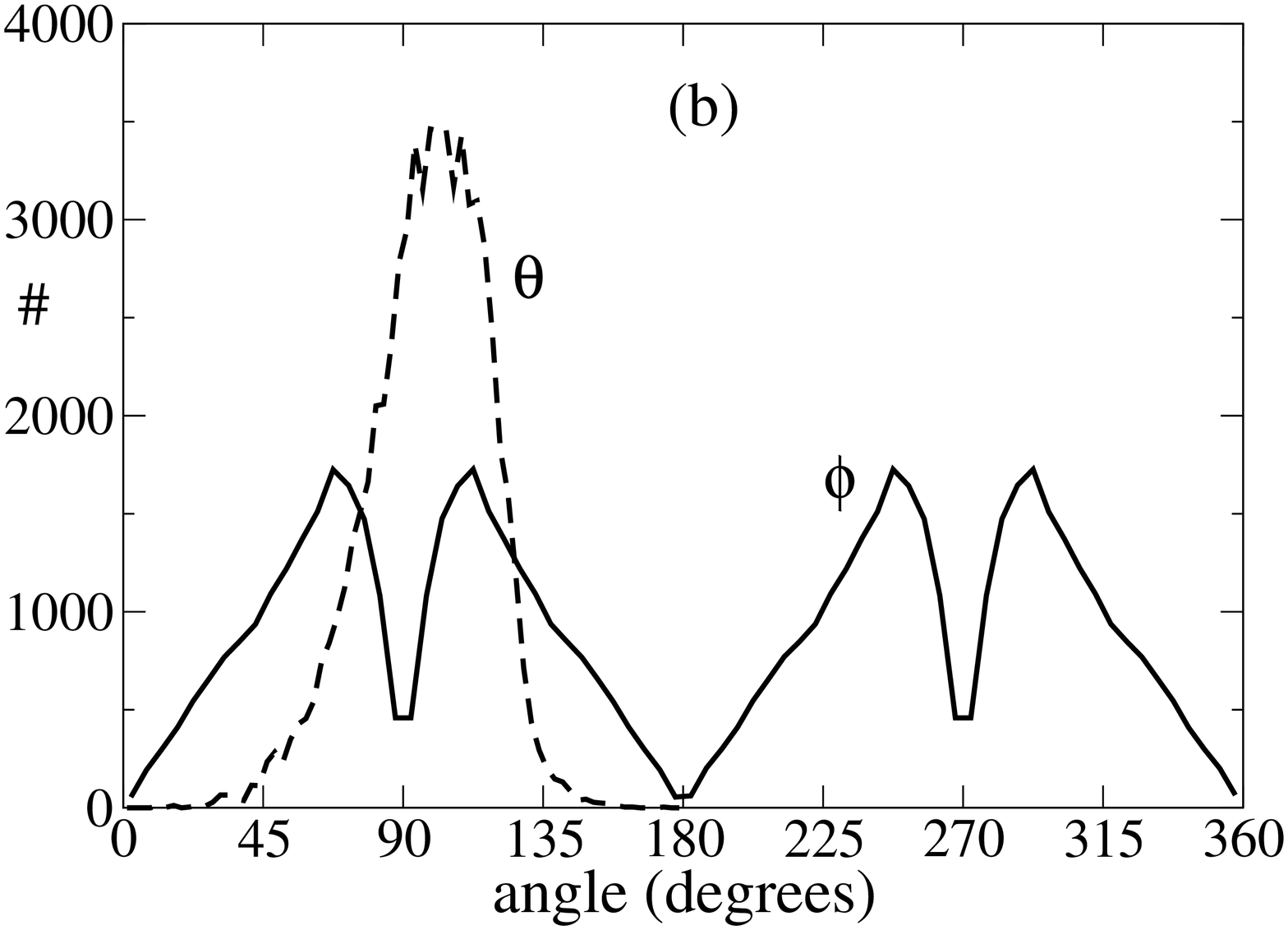,height=7.0cm,angle=-90}
%\end{center}

\caption{A scatter plot of $\theta$ vs. $\phi$ for the source
  population used in our simulations (a) as well as a distribution of
  those angles (b).  $\theta$ is the polar angle between the
  relative velocity vector of the potential impactor and the Earth's
  direction of motion and $\phi$ is an azimuthal location of the relative
  velocity \citep{val99}.  These figures show that our
  distribution of possible impactors is not isotropic, a key
  assumption in many cratering theories.  Only $\phi$
  values from $0 - 90$\degrees are shown in (a) corresponding to
  post-pericenter encounters at ascending node;  higher values of $\phi$ are
  simply mirror images of the scatter plot shown as all encounters
  have the same value of $\theta$ for a given object.} 
\label{fig:dirs}
\end{figure}

\begin{figure}[H]
\begin{center}
\epsfig{file=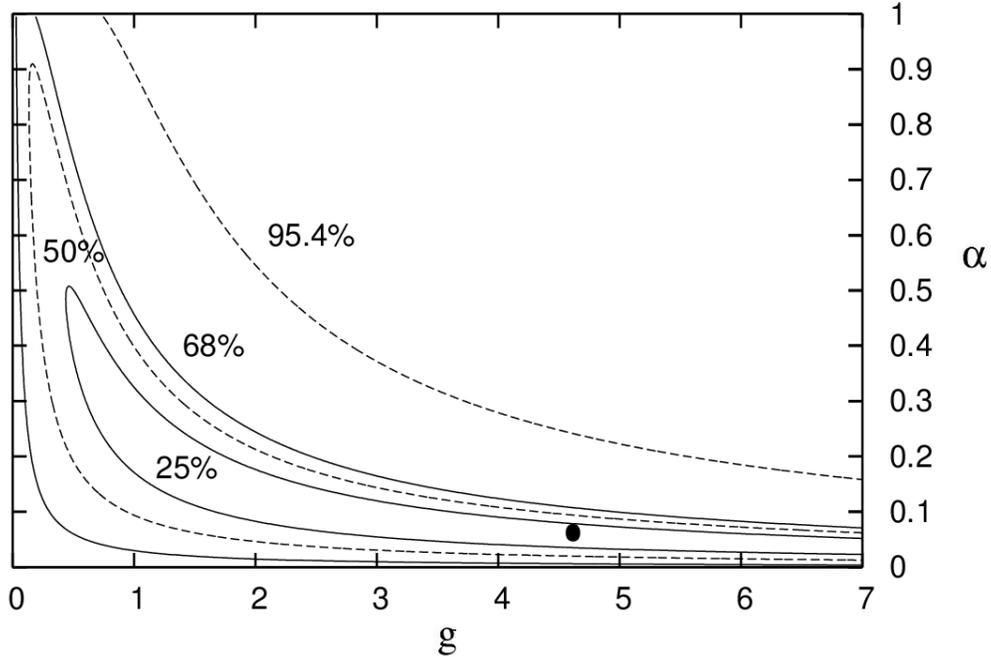,height=13.5cm,angle=-90}
\end{center}
\caption{A contour plot of log(likelihood) values showing the
  degeneracy present in the parameters $\alpha$ and $g$ from
  Eq.~\ref{eq:ass} using the rayed crater data from \citet{morota03}.
  Best fit values for the rayed crater data are shown as the dot at
  $\alpha = 0.063$ and $g = 4.62$.  This value for g gives the
  slope of the impactor mass distribution as $b = 1.87$, over 3 times the
  observationally determined value of 0.58.  However, a large
  range of $\alpha$ and $g$ values give acceptable fits to the data,
  making it impossible to measure the parameters of the projectile
  population from the crater distribution alone.}    
\label{fig:like}
\end{figure}

Although Eq.~\ref{eq:ass} may be fit to an observed crater
distribution, this does not necessarily provide a convenient measure
of the asymmetry.  Figure~\ref{fig:dirs} shows that for our impactor
population (see Sec.~\ref{sec-source}) the isotropic assumption fails,
so one should expect deviations from the functional form of
Eq.~\ref{eq:ass}.  Note that increasing either $\alpha$ or $g$ raises
the leading/trailing asymmetry, introducing a degeneracy in the
functional form, making it difficult to decouple the two when
determining information about \vinfty and the size distribution.  In
fact the observations actually permit a very large range of parameter
values.  Performing a maximum likelihood parameter determination using
the rayed crater data of \citet{morota03} yields Fig.~\ref{fig:like}.
Also, determining $\bar\Gamma$ with a high degree of precision from a
measured crater field is difficult.  To minimize these issues when
examining the entire lunar surface, we adopt the convention of
\citet{zetal01} by taking the ratio of those craters which fall within
30\degrees of the apex to those which are within 30\degrees of the
antapex.  This ratio forms a statistic known as the global measure of
the apex-antapex cratering asymmetry (GMAACA). 

Because of the Moon's small 1.0~km/s orbital velocity and standard
literature values of $\bar{v}_{\infty} \sim 10-16$ km/s
\citep{shoe83,chyba94}, one expects $\alpha \approx
1/16-1/10$ and thus a crater enhancement on the
leading hemisphere in the range $\sim 1.4-1.7$, in terms of the
GMAACA. Recent studies of young rayed craters
\citep{morota03,morota05} are found to be consistent with these
estimates. We will calculate the GMAACA that current NEO impactors
produce and compare with these values.

\newpage
\subsection{Lunar nearside/farside debate}
\label{sec-nearfar}
The concept of a nearside/farside asymmetry has garnered less
attention in the lunar cratering literature.  \citet{wiesel71}
discussed the idea that the Earth would act as a gravitational lens,
focusing incoming objects onto the nearside of the Moon.
However, the degree to which this lensing effect occurs is unclear.
The amplitude of nearside enhancement has been reported as insignificant
\citep{wiesel71}, a factor of two \citep{turk62,wood73}, and most
recently, a factor of four compared with the farside in a preliminary
analysis by \citet{fev05}.  In contrast, \citet{band73} used analytic
arguments to claim no a priori reason to expect such an asymmetry; the
Moon may be in a region of convergent or divergent flux because the
focal point of the lensed projectiles depends on the encounter
velocity of the objects and the Earth-Moon distance.  These latter
authors estimated that there should currently be a negligible
difference in the nearside/farside crater production rate.  We will
also investigate this issue as the initial conditions used in the
previous dynamical studies were somewhat artificial.

\section{Model and numerical methods}
\label{sec-MM}

\subsection{The source population}
\label{sec-source}

\begin{figure}[H]
\begin{center}
\includegraphics[height=9.5cm,angle=0]{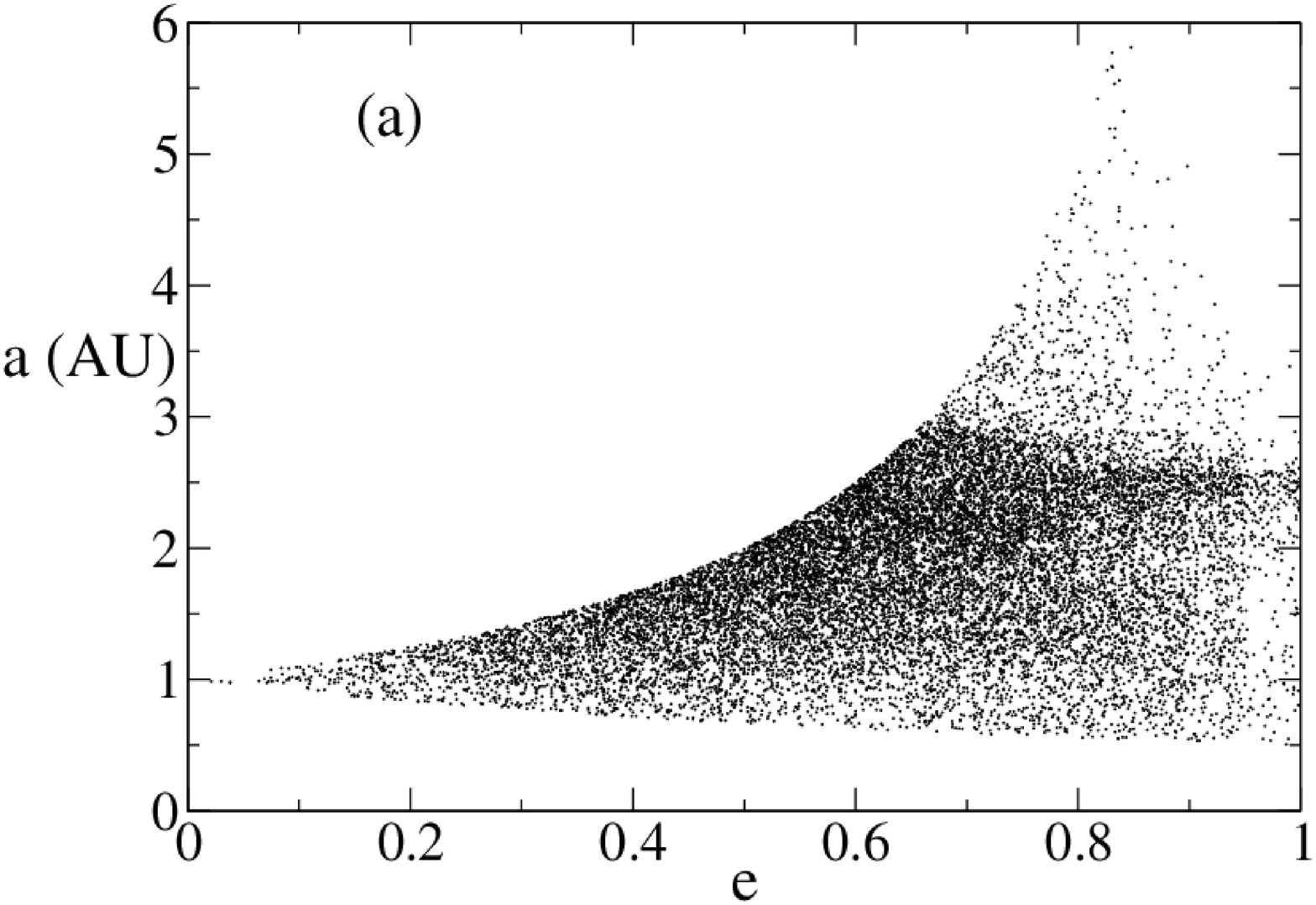}
\hspace{0.5cm} \includegraphics[height=9.5cm,angle=0]{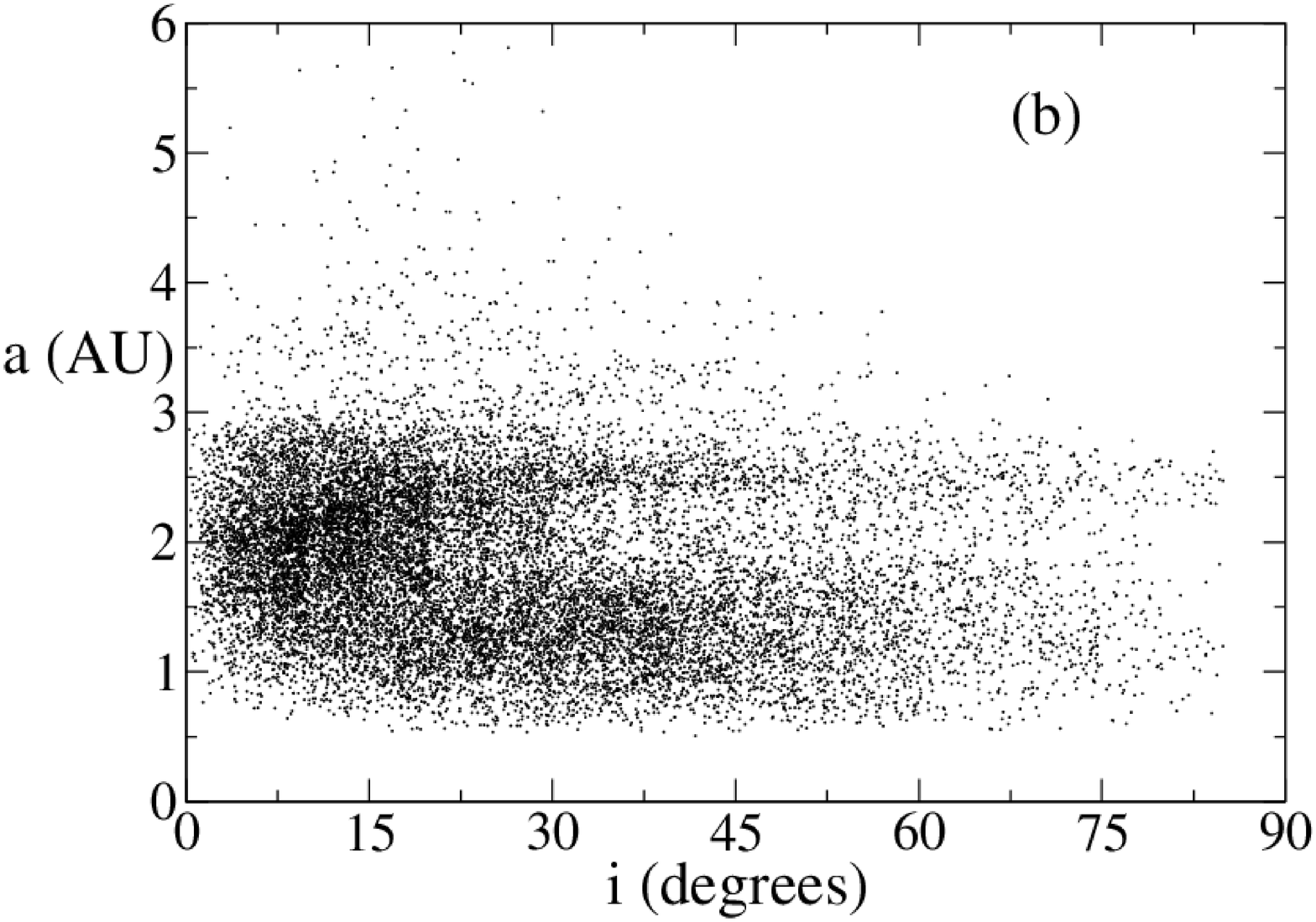}
\end{center} 
\caption{The orbital distribution of candidate impactors used in our
  simulations, based on the debiased near-Earth object (NEO) model
  of \citet{bottke02}.  We have used only objects whose perihelion $q$
  and aphelion $Q$ satisfy $q \leq 1$ AU and $Q \geq 1$ AU. (a)
  Semi-major axis $a$ versus eccentricity, $e$.  Note the increase in
  density near $a \sim 2.5$ AU due to the 3:1 orbital
  resonance with Jupiter.  (b) Semi-major axis versus inclination $i$
  with respect to the ecliptic.}  
\label{fig:aei} 
\end{figure}

The cratering asymmetry expected on the synchronously-rotating
Moon depends on the impactor speed distribution that is bombarding 
the satellite (the scalar \vinfty distribution and its directional
dependence).  Therefore, we must have a model of the small-body
population crossing Earth's orbit.  The potential impactors come from the
asteroid-dominated (\eg \citet{getal00}) near-Earth object (NEO)
population.  The orbital distribution of these objects is best modeled
by \citet{bottke02} who fit a linear combination of several main-belt
asteroid source regions and a Jupiter-family comet source region to
the Spacewatch telescope's NEO search results.  Since the detection
bias of the telescopic system was included, this yields a model of the
true NEO orbital-element distribution.  W.~Bottke (private
communication, 2005) has provided us with an orbital-element sampling
of their best fit distribution, which we have then restricted to 16307
orbits in the Earth-crossing region (Fig.~\ref{fig:aei}). This will be
our source population for the impactors which transit through the
Earth-Moon system.  We note the common occurrence of high eccentricity
$e$ and inclination $i$ orbits in the NEO sample, which result in high
values of \vinfty for the bombarding population. Although semimajor
axes $a$ in the $a$=2.0--2.5~AU region are densely populated (since
this is the semimajor axis range of the dominant main-belt sources),
impact probability is largest for orbits with perihelia $q=a(1-e)$
just below 1~AU or for aphelia $Q=a(1+e)$ just above 1~AU
\citep{morglad98}.  Since the $q\sim1$ population is much larger than
the $Q\sim1$ population, one might expect the former to dominate the
impactors (but see Sec.~\ref{sec-ampmsec}). 

\subsection{Setup of the flyby geometries}
\label{sec-setupgeo}

The orbital model provides only the $(a,e,i)$ distribution of
the NEO population and we expect that the 1.7\% eccentricity
of the Earth's orbit is a small correction to the impactor
distribution, so for what follows we have modeled the Earth's
heliocentric orbit as perfectly circular.  As a result we can,
without loss of generality, take all encounters to occur at (1~AU,0,0)
in heliocentric coordinates, where we have lost knowledge of the day
of the year of the encounters (although we can easily average over the
year by {\it post-facto} selecting a random azimuth for the Earth's
spin pole at the time of a projectile's arrival at the top of the
atmosphere).  With this restriction, each Earth-crossing orbit can
have an encounter in one of four geometries depending on the argument
of pericenter $\omega$ and the true anomaly $f$, which must satisfy:   
\begin{equation}
r = 1~\mathrm{AU} = \frac{ a(1-e^2) }{1 + e \cos  f } \; \; \; .
\end{equation}
By construction, the encounter must occur at either the ascending
or descending node along the $x$-axis, and so the longitude of 
ascending node is $\Omega$=0 or $\pi$.  Taking $f$ and $\omega$ in
$[0,2\pi)$, for ascending encounters, $f=2\pi-\omega$ for either
  encounters with pericenters above ($\omega=[0,\pi)$) or below
    ($\omega=[\pi,2\pi)$) the ecliptic.  If the encounter occurs at
      the descending node, $f = \pi - \omega$ for post-pericenter
      encounters and $f = 3\pi - \omega$ for pre-pericenter
      encounters.  With this in mind, we effectively quadruple the
      number of initial conditions to 65228. 

With the longitude of ascending node fixed (ie. $\Omega = 0
\mathrm{\ or\ }\pi$), we then construct a plethora of incoming initial
conditions based on the orbital elements from the debiased NEO model
converted to Cartesian coordinates.  For each initial orbit, all four
of the encounter geometries are equally likely.  We randomly choose a
particle for a flyby based on its encounter probability with the Earth
as judged by an \"Opik collision probability calculation
\citep{dones99}.  Gravitational focusing by the Earth was {\it not}
included in the encounter probability estimate as an increased
frequency of Earth deliveries will occur naturally during the flyby
phase if the Earth's gravity is important.   

Both Earth and a test particle (TP) are then placed at the nodal
intersection and moved backwards on their respective orbits
until the separation between the NEO and the Earth is 0.02~AU.  At
this point, we create a disk of $10^5$ non-interacting test particles,
meant to represent potentially-impacting asteroids or comets, centered on the 
chosen orbit.  A short numerical integration is run for each of the
65228 initial conditions and one TP trajectory that results in an
arrival at the Earth in each flyby is placed into a table of new
initial conditions. This procedure is performed because during the
``backup phase'' to separate the Earth and TP by 0.02~AU,
gravitational focusing was not accounted for.  The omission could
result in the particle missing the Earth in a forward integration,
as the Earth's gravity would be present, modifying the chosen
trajectory.  This gives us a final set of 65228 initial trajectories
that strike the Earth when a forward integration is performed.  For
convenience we then convert from heliocentric coordinates to a
geocentric frame of reference.   

%This separation of two Earth Hill Sphere radii between the Earth and
%the TP orbit has been chosen to avoid shearing of the TP disk which
%causes artifacts to appear in the spatial distribution of deliveries.

\begin{figure}[H]
\begin{center}
\epsfig{file=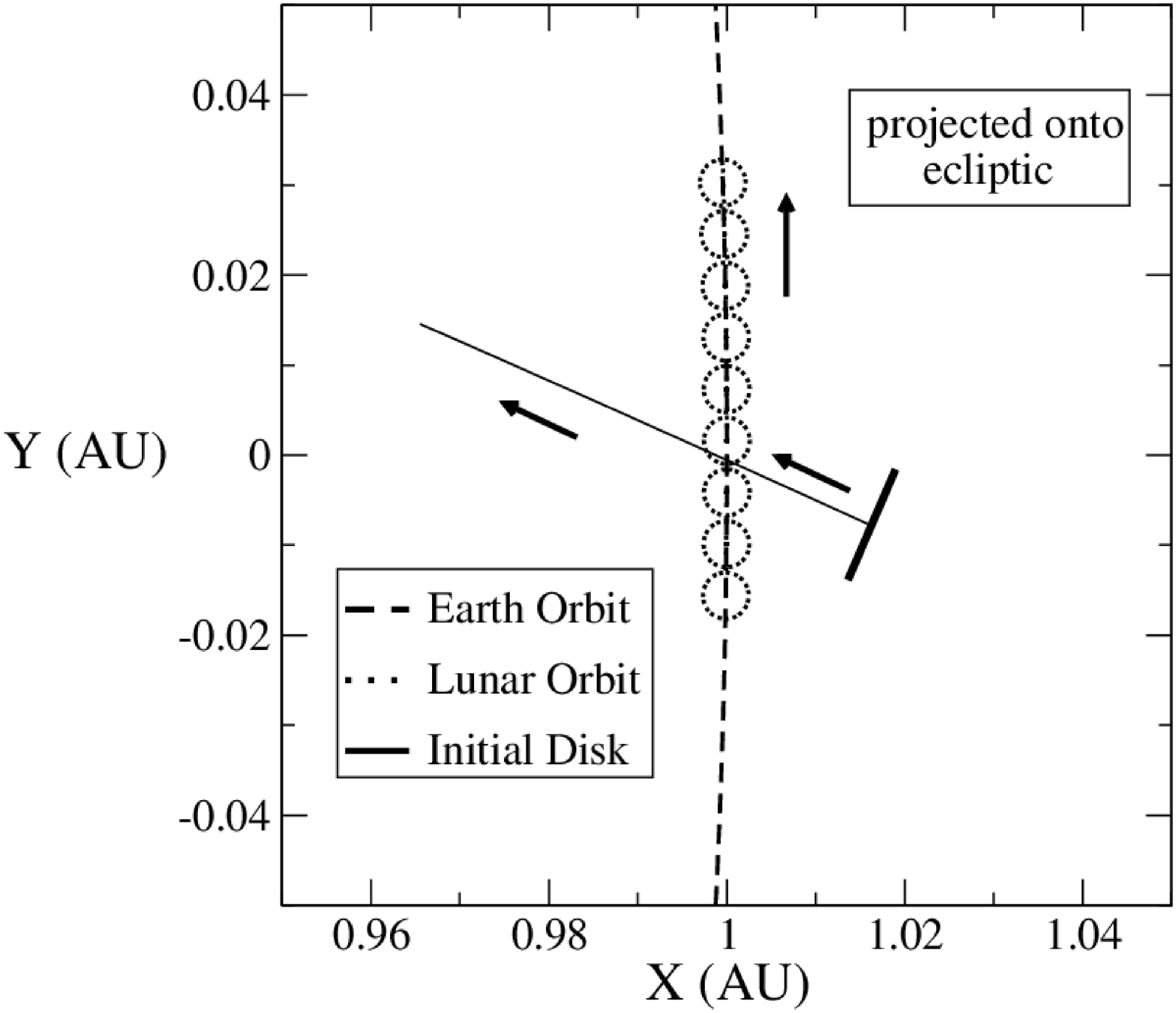,height=12.5cm,angle=0}
\end{center} 
\caption{A schematic of one 3-dimensional flyby projected onto the
  ecliptic in a heliocentric coordinate frame.  A disk (thick, solid
  line) of $10^5$ test particles, all with identical initial
  velocities, is created around a central trajectory based on the NEO
  source population of \citet{bottke02}.  The disk is 2.5 lunar
  orbital radii in radius and encompasses the entire lunar orbit for
  any of our possible initial conditions.  These test particles are
  then integrated forward along with the Earth, Moon, and Sun and have
  an encounter with the Earth-Moon system near (1~AU,0,0).  All
  particles are followed for three times the amount of time required
  for the farthest test particle to reach the Earth.}  
\label{fig:setup} 
\end{figure}

For our simulations, one of the new initial conditions is randomly
chosen based on a newly calculated encounter probability with the
Earth-Moon system.  A new disk, centered on the chosen trajectory, is
randomly populated with $10^5$ test particles, all given identical
initial velocities.  The radius of the disk is chosen to be 2.5 lunar
orbital radii as testing showed that for all of our initial conditions,
a disk of this size spans the entire lunar orbit as it passes the
Earth's position.  This was done to ensure the crater distribution had
no dependence on the lunar orbital phase.  Figure~\ref{fig:setup} shows
a schematic of the simulations.

\subsection{Integration method} 

The orbital trajectories of the massive bodies (Earth, Moon and
Sun) and the TPs are integrated using a modified version of
swift-rmvs3 \citep{levdun94}.  The length units are in AU 
and time units are in years.  The initial eccentricity and inclination
for both the Earth and Moon are set to zero for the majority of our
simulations.  In one run we introduce the current 5.15\degrees
inclination of the Moon, keeping Earth's $e$ and $i$ as before.  

Both Earth and Moon are assumed to be perfect spheres.  Factors that vary
in the simulations include the number of flybys, where a flyby is
defined as a disk of $10^5$ test particles passing through the
Earth-Moon system, and the Earth-Moon distance, $R_{EM}$. 

We have the Earth as the central body with the Moon
acting as an orbiting planet and the Sun as an external perturber.
The base time step in the simulations is four hours.  This is large
enough to not be time prohibitive, but small enough such that the
integrator can follow the lunar encounters precisely.  During the
course of the integration, the integrator logs the positions and
velocities of all bodies of interest if there is a pericenter passage
within the radius of the Moon or Earth.  

We then use this log as input for a backwards integration using a
$6^{th}$-order explicit symplectic algorithm \citep{gladsimp91} to
precisely determine the latitude and longitude of the particle's
impact location.  This method, along with iterative time steps,
enables us to find impact locations to within 5~km of the surface of
the Moon and within $< 500$~m of the top of Earth's atmosphere.

This project is computationally intensive.  For each simulation,
the Earth, Moon and Sun are included as well as $10^5$ test
particles.  Each of the 65228 initial conditions are used multiple
times to improve statistics and in each case the TP locations in the
disk are randomly distributed.  Thus more than $2 \times 10^{12}$
test particles are integrated which results in tens of millions of
terrestrial deliveries and hundreds of thousands of lunar ones.

\section{Delivery to Earth}
\label{sec-ED}

Before turning to the Moon, we examine our numerical results to
determine the spatial distribution of Earth arrivals.  Our simulations 
with the Moon at the current orbital distance of $60R_\oplus$
yielded 21,998,427 terrestrial impacts.  Since the crater record on the
Earth is difficult to interpret due to geological processes, we choose
to examine our results in terms of fireball and meteorite records as
the impactors strike the top of the atmosphere.  We are thus assuming
that the meteorite dropping bodies have a pre-atmospheric orbital
distribution similar to the NEOs (but see Sec.~\ref{sec-ampmsec}).

\begin{figure}[H]
\begin{center}
\epsfig{file=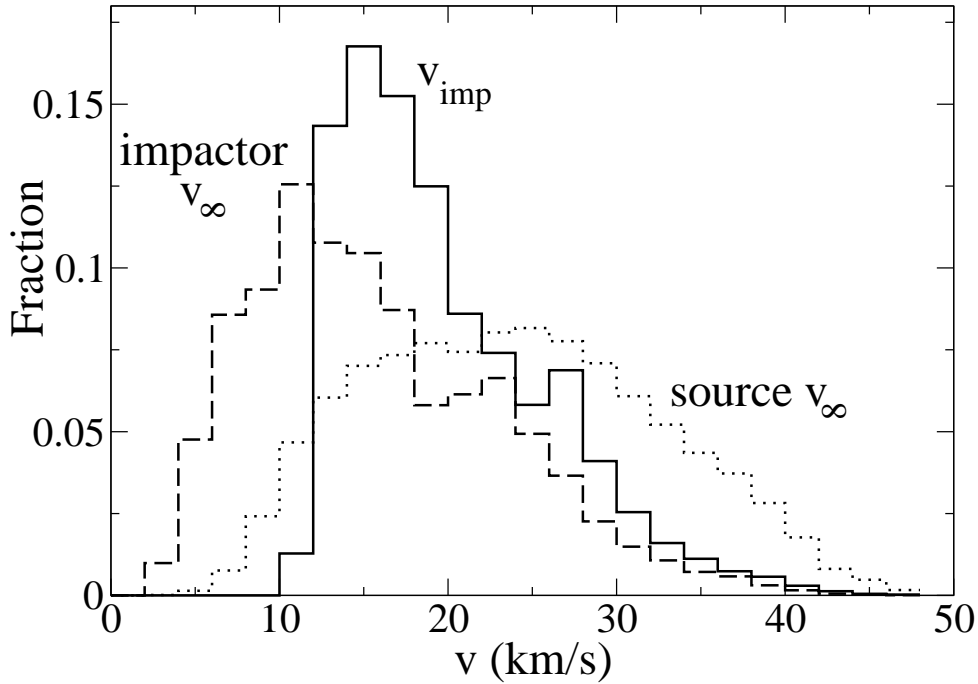,height=14.5cm,angle=-90}
\end{center} 
\caption{Velocity distributions for the Earth deliveries from our
  simulations (impactor \vinfty and $v_{imp}$) as well as the sampled
  NEO population (source \vinfty).  Note that the average impact
  velocity is $\sim$~20 km/s, higher than the often-quoted value of
  $17$ km/s.  One can observe the large effect of the Earth's
  gravitational well for \vinfty values $\lesssim 10$ km/s.  Here the
  encounter speeds are ``pumped up'' to values slightly higher than
  the Earth's escape velocity of 11.2 km/s.  From the \vinfty
  distribution of $all$ possible impactors it is clear that the
  impacts are dominated by low (\vinfty$\lesssim 20$ km/s) objects.}  
\label{fig:evhist}  
\end{figure}

Figure~\ref{fig:evhist} shows velocity distributions for the Earth
arrivals from our simulations.  The delivery speeds should be
interpreted as ``top of the atmosphere'' velocities, $v_{imp}$.  The
deliveries are dominated by low \vinfty objects which the Earth's
gravitational well has focused and sped up, creating a peak in the
distribution at a speed of $\sim 15$ km/s.  The average speed an
impactor has at the top of the atmosphere is $\sim 20$ km/s, slightly
higher than the often quoted value of 17 km/s \citep{chyba94}.  As
well, if we compare our Fig.~\ref{fig:evhist} to the fireball data in
Morbidelli and Gladman (1998, Fig.~8a), we see a general similarity in
the shapes of the $v_{imp}$ histograms. 

\vspace{-1.0cm}

\newpage
\subsection{Latitude distribution of delivered objects}
\begin{figure}[H]
\begin{center}
\epsfig{file=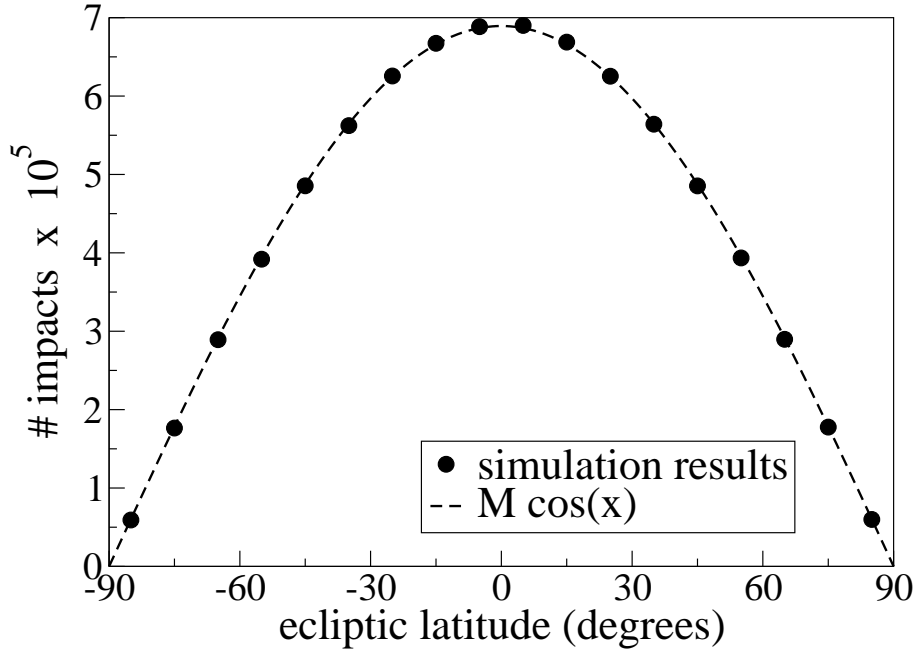,height=13.5cm,angle=-90}
\end{center} 
\caption{A latitude distribution of projectile deliveries to
  the Earth from one set of simulations.  This figure has not
  accounted for the spin obliquity of the planet (hence ``ecliptic
  latitude'').  The points represent the total number of impacts in
  $10^\circ$ ecliptic latitude bins and the dashed curve is a cosine
  multiplied by an arbitrary constant, $M$.  The error bars for the
  simulation results are smaller than the points.}
\label{fig:elat} 
\end{figure}

For a uniform spatial distribution of deliveries, one expects the
number of arrivals to vary as the cosine of the latitude due to the
smaller surface area at higher latitudes.  Figure~\ref{fig:elat} shows
that to an excellent approximation, the Earth is uniformly struck by
impactors (in latitude).  To account for the area in each latitude
bin, divide by: 
\begin{equation}
A_{bin} = 2\pi R_{\oplus}^2(\cos\theta_1 - \cos\theta_2),
\end{equation}
where the co-latitudes $\theta_1 > \theta_2$ are measured north from
the south pole.  

\vspace{-1.0cm}

\begin{figure}[H]
\begin{center}
\epsfig{file=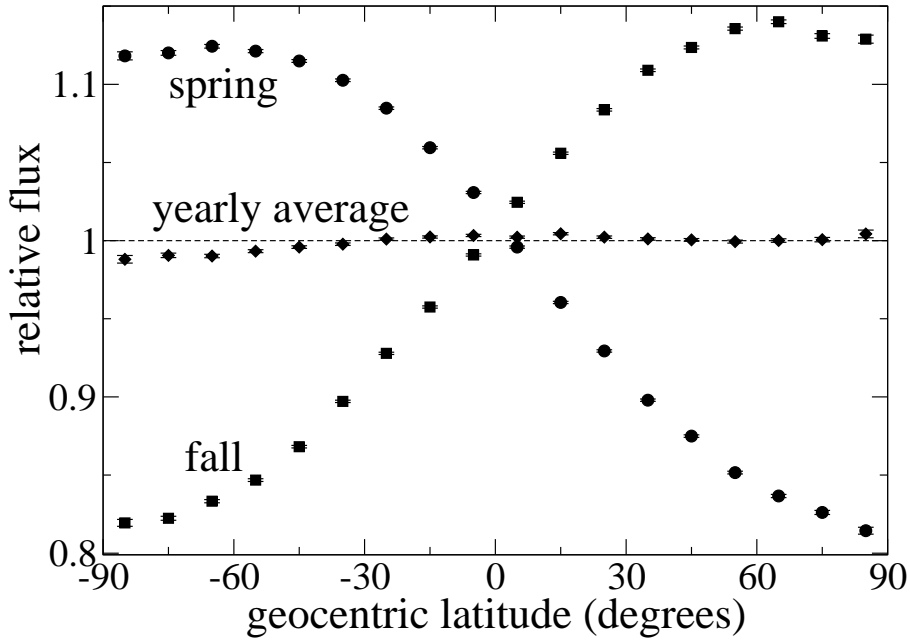,height=13.5cm,angle=-90}
\end{center} 
\caption{The geocentric latitude distribution, relative to the average
  spatial density of impacts, at different times of the year.  The
  circles represent northern hemisphere spring, when the Earth's spin
  axis (as viewed from above) is tilted away from the planet's
  direction of motion.  The near mirror-image can be seen in northern
  hemisphere fall (squares) when the spin axis is tilted towards the
  Earth's direction of motion.  Averaged over a full year (diamonds),
  the deviation from the expected flat distribution is minimal.}   
\label{fig:elat_var} 
\end{figure}

\noindent To correct for the spin obliquity of the
Earth, we then choose a random day of fall, and thus a random azimuth for
the Earth's spin pole.  This then gives us a geocentric latitude and
longitude for each simulated delivery.  Figure~\ref{fig:elat_var} shows
the spatial density versus geocentric latitude.  For the long-term
average, we see a nearly uniform distribution of arrivals.  If we
restrict to azimuths corresponding to northern hemisphere spring or
northern hemisphere autumn, we see the well-known seasonal variation
\citep{xmas82,renkno89} in the fireball flux of roughly 15\%
amplitude (Fig.~\ref{fig:elat_var}).  As a measure of the asymmetry
between the poles and equator, we take the ratio between
polar (within 30\degrees of the poles) and equitorial (a 30\degrees
band centered on the equator) arrival densities.  We find that when all
terrestrial arrivals are considered, the poles receive the same flux
of impactors as the equator to $< 1$\% ($0.992 \pm 0.001)$.  We
believe the uncertainty caused by our finite sampling of the orbital
distribution is on this level and thus our results are consistent with
uniform coverage. 

\vspace{-1.0cm}

\begin{figure}[H]
\begin{center}
\epsfig{file=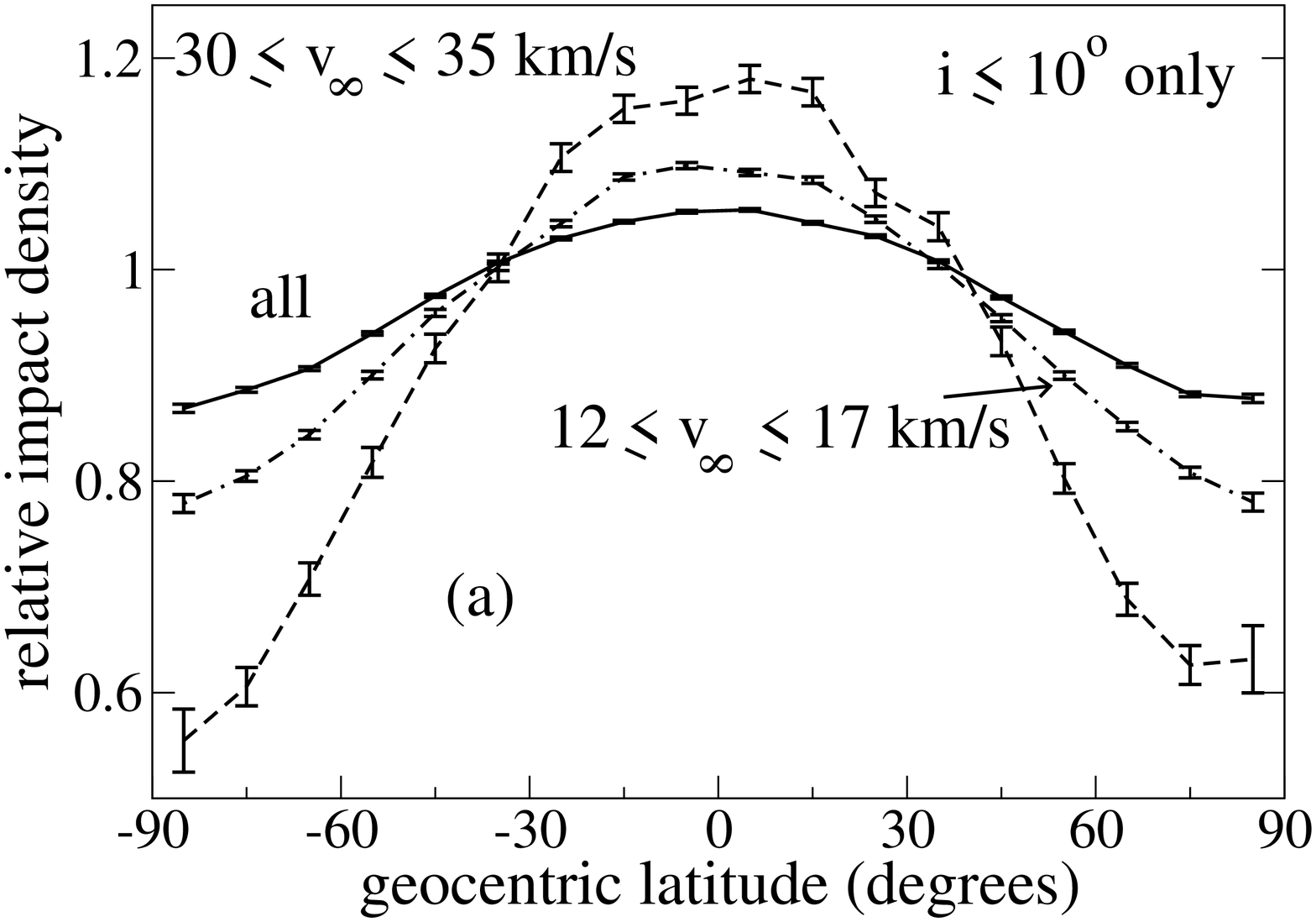,height=8.0cm,angle=-90}\hspace{-0.5cm}
\epsfig{file=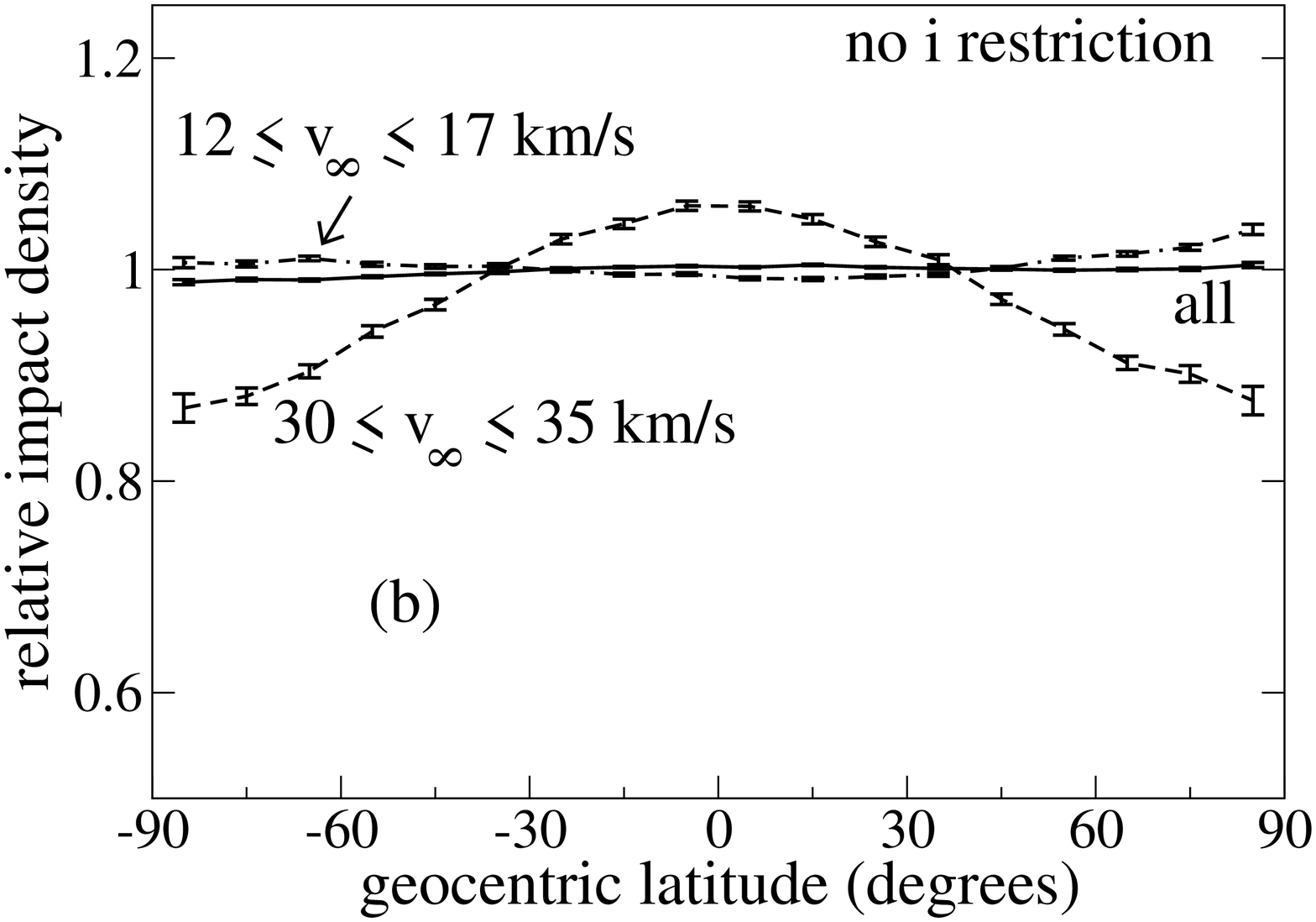,height=8.0cm,angle=-90}
\end{center}
\caption{The terrestrial impact density versus geocentric latitude (a) for
  different ranges of encounter velocity \vinfty and restricted to objects
  with inclination $\leq 10$\degrees and (b) the same with no
  inclination restriction.  For lower \vinfty objects, there
  is less latitudinal variation because the Earth's gravitational
  field bends incoming trajectories towards higher latitudes.  This
  effect is muted as the velocities become larger - the trajectories
  move in straighter lines.  For these faster objects, the available impact
  area varies as a cosine so the shape of the distribution for high
  \vinfty impactors is expected to do the same.  The deviations from a
  cosine are due to the fact the objects have moderate inclinations and
  non-infinite velocities.  Note that (b) shows the realistic
  case of all encounter speeds and inclinations; there is very little
  latitudinal variation.}  
\label{fig:fevcomp1}
\end{figure}

\newpage
To better understand the details of latitudinal asymmetries, we consider the
effect of approach velocity, \vinfty.  When our projectile deliveries
are filtered such that $v_{cut1} \leq v_{\infty} \leq v_{cut2}$, and
restricted to impactors with $i \leq 10$\degrees we find results
similar to Le Veuvre and Wieczorek (2006, our Fig.~\ref{fig:fevcomp1}a 
can be compared to their Fig.~2).  Their $N_g$ parameter is analogous to
our \vinfty cuts.  Small encounter velocities produce
more uniform coverage because the trajectories are bent towards the
poles.  Higher encounter velocity trajectories are not effected to as
great a degree and tend to move in nearly straight lines, leading to a
distribution which tends to a cosine-like curve.  Obviously when the
$i$ restriction is lifted, the poles receive a higher flux which mutes
the amplitude of the variation (Fig~\ref{fig:fevcomp1}b).

\vspace{-1.0cm}

The preliminary results of \citet{fev06} show a 30\% ecliptic
latitudinal variation in terrestrial projectile deliveries.  Though we
are able to produce results similar to theirs under various
restrictions (Fig.~\ref{fig:fevcomp1}a), the inclusion of the Earth's
spin obliquity is necessary to accurately reflect reality.  With this 
inclusion, we find a latitude distribution which is very nearly
uniform (Fig.~\ref{fig:fevcomp1}b). 

\subsection{AM/PM asymmetry}
\label{sec-ampmsec}

The ecliptic latitudes and longitudes for the terrestrial arrivals
were converted into local times via a straightforward method.  As
stated in the previous section, the Earth's spin-pole azimuth is
chosen at random to represent any day of the year and then the arrival
location is transformed to these geocentric coordinates.  The location
relative to the sub-solar direction gives the local time.   

\begin{figure}[H]
\begin{center}
\epsfig{file=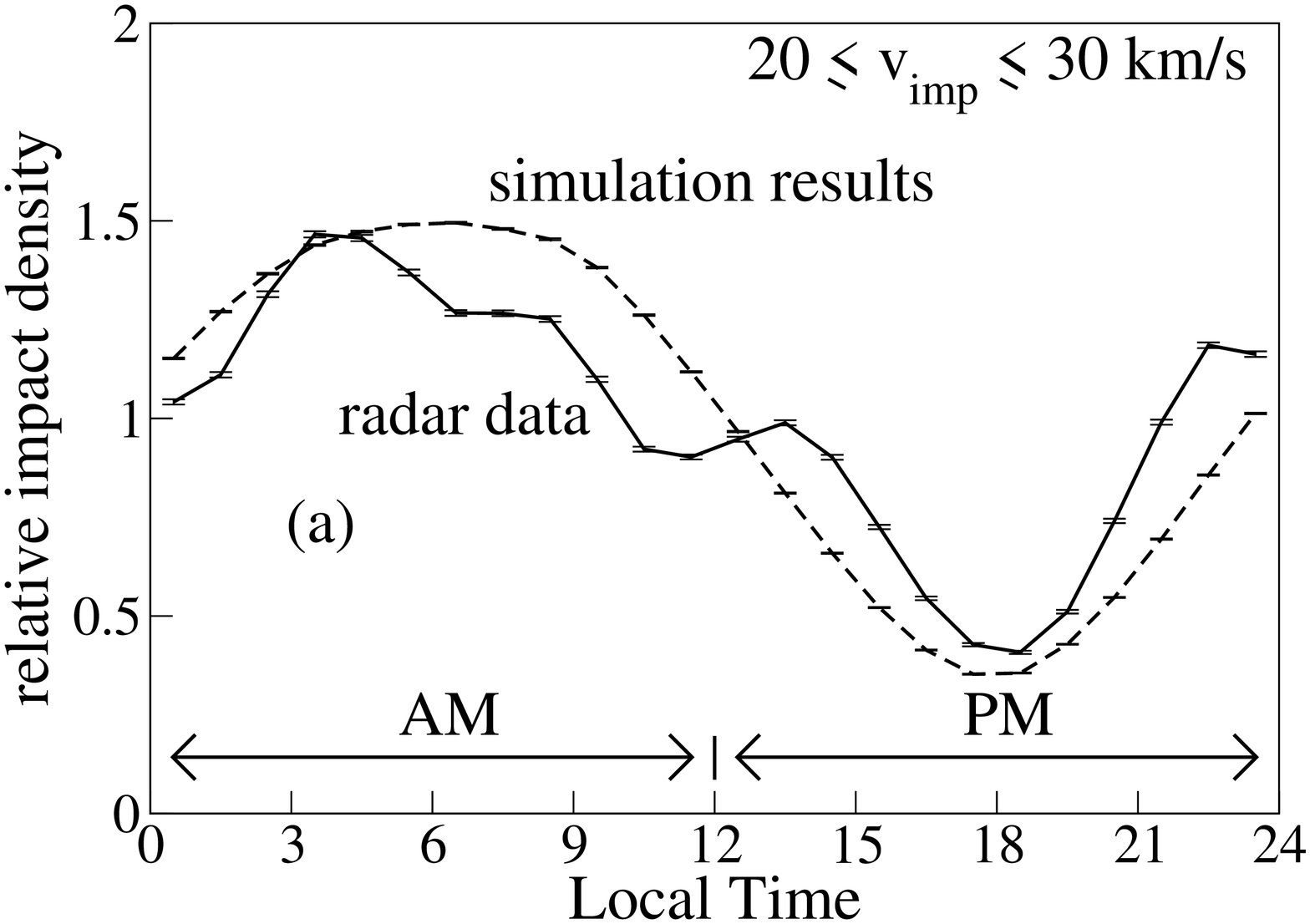,height=8.5cm,angle=-90}\hspace{-0.75cm}
\epsfig{file=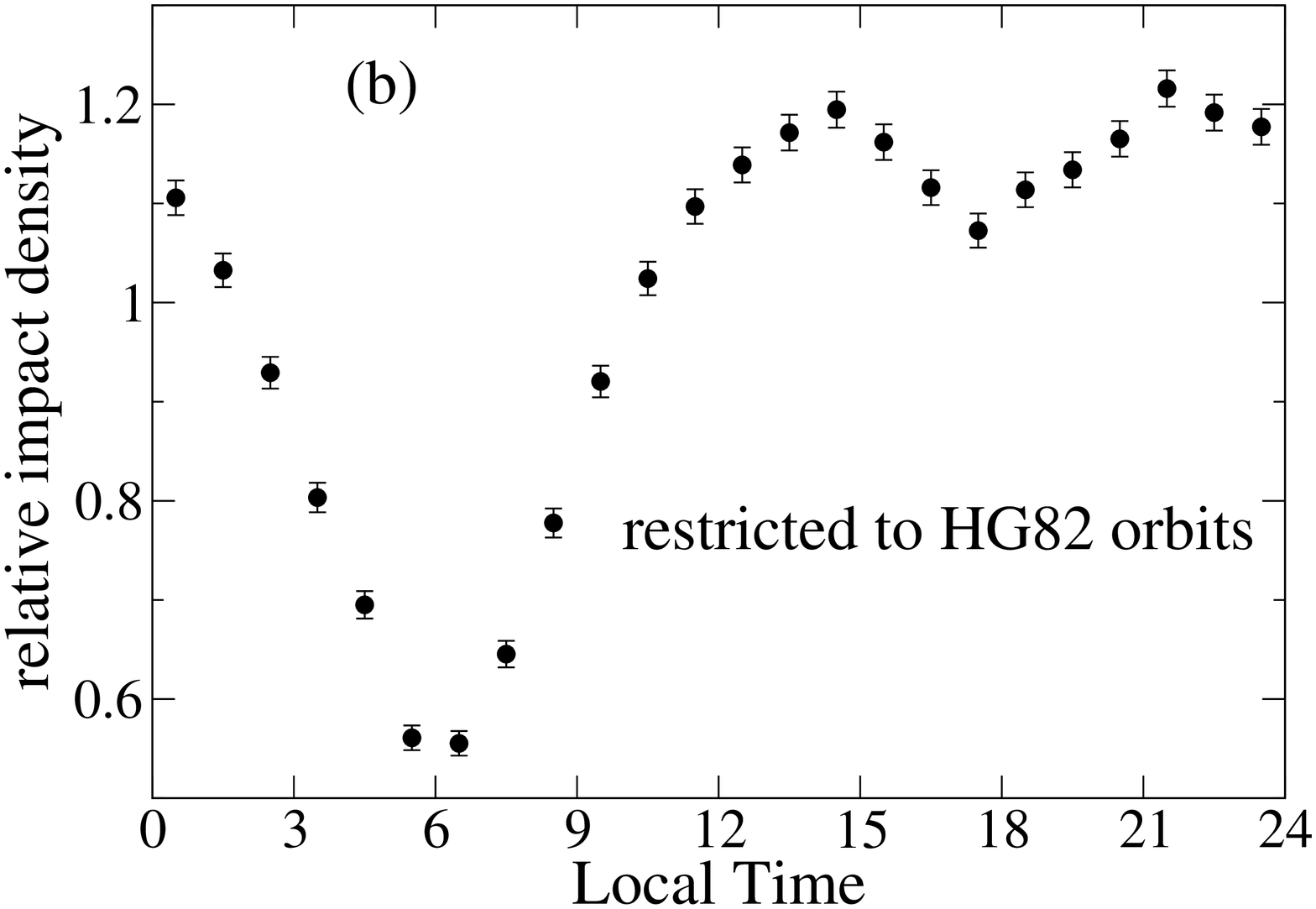,height=8.5cm,angle=-90}
\end{center} 
\caption{The local time distribution of arrivals at the top of Earth's
  atmosphere for both our simulations and radar data (a).  The
  simulation deliveries have been corrected for the Earth's spin
  obliquity.  Clearly there is a large AM excess in the arrivals
  which is counter to the meteorite fall data which show a
  PM enhancement.  Though a velocity restriction is used (see text),
  other cuts yield the same general shape for both radar data and
  simulation results.  (b) The local time distribution of Earth
  deliveries when restricted to orbits similar to the ones used in
  \citet{xmas82}.  This distribution matches well with the time of
  falls for chondrites.}  
\label{fig:ampm}
\end{figure}

We looked for a PM excess in our simulation results.  However, as
evident in Fig.~\ref{fig:ampm}a we see the opposite effect.  Previous
modeling work produced a near mirror image (reflected through noon) of
our result (see our Fig.~\ref{fig:ampm}b and Fig.~3 of \citet{xmas82}),
but their impactor orbital distribution was very different from the
debiased NEO model; they chose a small set of orbits with perihelion
$q$ in the range $0.62 \leq q \leq 0.99$~AU with semimajor axis $a$
obeying $1.3 \leq a \leq 3.2$~AU.  If we restrict our simulation
results to approximately the same impactor distribution (by taking
only those objects having $a$ and $q$ within $\pm\ 0.01$~AU of the
entries in Table 1 of \citet{xmas82}), we obtain Fig.~\ref{fig:ampm}b
which is very similar to their result. 

\vspace{-1.0cm}

\begin{figure}[H]
\begin{center}
\epsfig{file=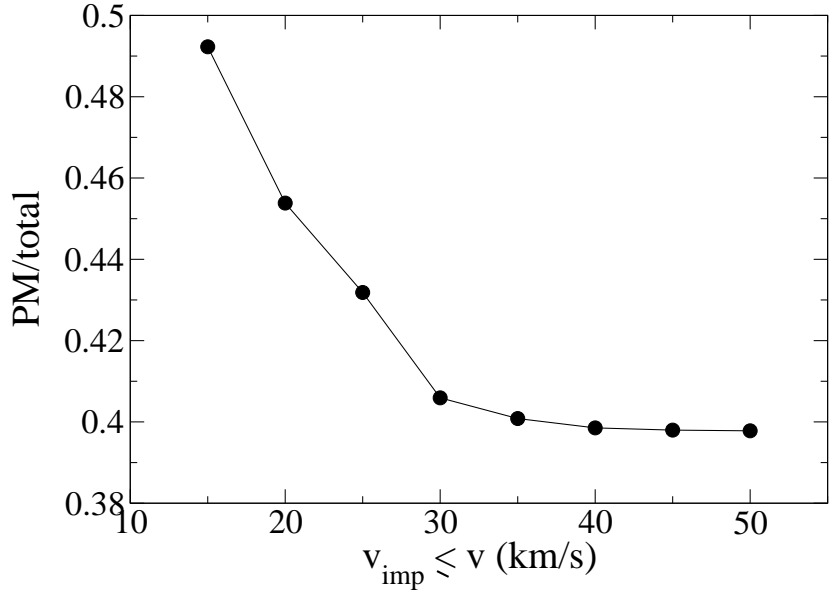, height=12.0cm, angle=-90}
\end{center}
\caption{  The fraction of Earth arrivals on the PM hemisphere, as a
  function of various cutoff speeds in the incoming flux. For the
  entire population, the PM ratio is 40\%. Applying more and more
  stringent upper bounds pushes the PM ratio towards 50\%, but for
  reasons discussed in the text we do not believe that the NEO orbital
  distribution is the same as that of meteorite-dropping fireballs,
  and thus the AM excess we find (which is not exhibited by the
  fireball data as a whole) does not create a contradiction with the
  available data. }     
\label{fig:pmtot}
\end{figure}

Why is there this discrepancy with our unrestricted case?  The
origin is not one of method but rather of starting conditions. The
debiased NEO model we use is much more comprehensive than the orbits
used by \citet{xmas82}, where only orbits assumed to represent
then-current fireballs were included.  The real Earth-crossing
population contains a larger fraction of high-speed orbits, which
produce a smaller fraction of PM falls than the shallow
Earth-crossers.  In fact, our simulated arrivals always show an AM
excess (Fig.~\ref{fig:pmtot}), even if we apply an upper speed bound
(in an attempt to mimic a condition for meteoroid survivability,
requiring speeds of less than 20--30~km/s at the top of the
atmosphere).  

\vspace{-1.0cm}

The apparent conflict with the meteorite data should not be too
surprising, as \citet{morglad98} argued that that orbital distribution
of the 0.1--1~m-scale meteoroids that drop chondritic fireballs
must be different than that of the Near-Earth Objects.  They showed
that to match the radiant and orbital distributions determined by the
fireball camera networks and to also match a PM excess, these
sub-meter sized bodies must suffer strong collisional degradation as
they journey from the asteroid belt, with a collisional half life
consistent with what one would expect for decimeter-scale objects;
this produces a match  with the fireball semimajor axis distribution,
which is dominated by the $a>1.5$~AU orbits.  In contrast, our
simulations show NEO arrivals are much more dominated by $a\sim1$~AU
objects.  We posit this is further evidence that the source region for
the majority of the meteorites (the chondrites) is the main belt and
not near-Earth space; to use the terminology of \citet{morglad98}, the
`immediate precursor bodies', in which the meteoroids were located
just before being liberated and starting to accumulate cosmic-ray
exposure, are {\it not} near-Earth objects but must be in the main
belt.   

As a consistency check, we compare our fall time distribution to 
radar observations \citep{jones05} of meteoroids arriving at the top
of Earth's atmosphere.  Figure~\ref{fig:ampm}a shows that the fall
time distribution obtained with our simulations is similar to the flux
of radar-observed meteors when restricted to the same
top-of-the-atmosphere speed range of $20 \leq v_{imp} \leq 30$ km/s
(though other cuts yield similar results).  The typical
pre-atmospheric masses of the particles producing the radar meteors is
in the micro- to milli-gram range (P. Brown, private communication
2006).  The velocity range chosen for Fig.~\ref{fig:ampm}a removes the
low-speed fragments which have reached Earth-crossing by radiation
forces (unlike the NEOs of the Bottke model, whose orbital
distribution is set by gravitational scatterings with the terrestrial
planets) and also removes the high-speed cometary component.  

The match we find permits the hypothesis that the majority
of the milligram-particle flux on orbits with these encounter speeds is
actually dust that is liberated continuously from NEOs, in stark
contrast with the decimeter-scale meteoroids, who must be recently
derived from a main-belt source.

\section{Lunar bombardment}
\label{sec-LB}

\begin{figure}[H]
\begin{center}
\epsfig{file=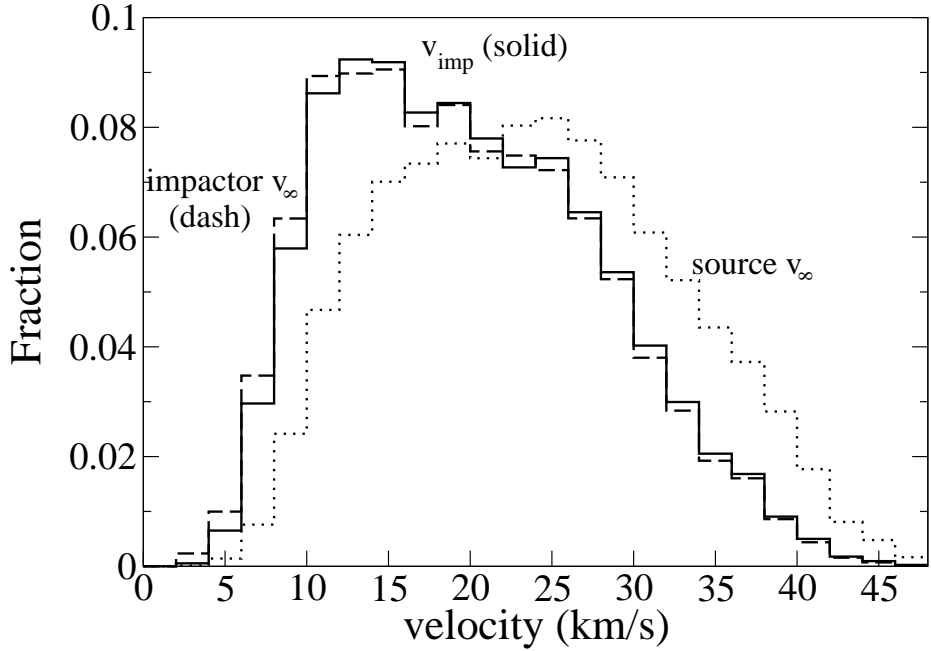,height=13.5cm,angle=-90}
\end{center} 
\caption{Velocity distributions for the lunar impacts from our
  simulations (impactor \vinfty and $v_{imp}$) and the sampled NEO
  source population (source \vinfty).  The average impact velocity, is
  $\sim$~20     km/s.  The curve showing the \vinfty of objects
  striking the Moon closely matches the \vimp distribution as one
  expects due to the small degree at which the moon's gravity well
  ``speeds up'' low \vinfty objects.  The difference between the
  $v_{imp}$ and source \vinfty distributions shows that gravitational
  focusing favours low speed (\vinfty$\lesssim 20$ km/s) objects.
  These simulation results were for the current Earth-Moon distance
  with the Moon orbiting in the plane of the ecliptic.}   
\label{fig:velhist} 
\end{figure}

Figure~\ref{fig:velhist} shows the lunar impact speed distribution from
our simulations.  Because the Moon's orbital and escape speeds
(1.02 and 2.38 km/s respectively) are both small compared to typical
\vinfty encounter speeds, the impacts are not as biased towards
smaller speeds as for the Earth.  As one would expect, since
$v_{imp}^2 =  v_{\infty}^2 + v_{esc}^2$, the $v_{imp}$ and source
\vinfty distributions are quite similar.  The small difference arises
from gravitational focusing by the Moon, which increases the speeds of
the low \vinfty population.  We compute the average impact speed
for NEOs striking the moon to be 20 km/s.  This is higher than the
often quoted lunar impact velocity of 12-17~km/s
\citep{chyba94,strom05}.  These lower velocities have been derived
using only the known NEOs and are therefore biased towards objects
whose encounter speeds are lower (which are more often observed in
telescopic surveys).   

The debiased NEO distribution we use has a full suite of high-speed
impactors; more than half the impactors are moving faster
than $v_{med}$ = 19.3 km/s when they hit the Moon.  This has serious
implications for the calculated projectile diameters that created
lunar craters since the higher speeds we calculate mean that typical
impactor diameters are smaller than previously derived.  Strictly
speaking, our results apply only when the {\it current} NEO orbital
distribution is valid, which has likely been true since the post-mare
era.  However, the generically-higher lunar impact speeds we find are
likely true in most cases for realistic orbital distributions, and
thus the size-frequency distribution of the impactors must be shifted
to somewhat smaller sizes in trying to find matches between the lunar 
crater distribution and the NEA size distribution (see Strom et al
2005 for a recent example).

The reader may be surprised to see that the average lunar impact speed
is essentially the same as the average arrival speed at Earth despite
the acceleration impactors receive as they fall into the deep gravity
well of our planet.  This (potentially counter-intuitive) result can
be understood once one realizes that Earth's impact speed distribution
is heavily weighted towards low \vinfty values by the Safronov factor
$ (1 + v_{esc}^2 / v_{\infty}^2 )$.  For Earth this so heavily
enhances the low encounter speed impactors that the average impact
speed actually drops to essentially the same as that of the Moon
(which does not gravitationally focus the low $v_{\infty}$ nearly as
well).  While the Earth's greater capture cross-section ensures a much
larger total flux, the average energy delivered per impact will be
similar for both the Moon and Earth. 

\vspace{-0.5cm}
\subsection{From impacts to craters}
\begin{figure}[H]
\begin{center}
\epsfig{file=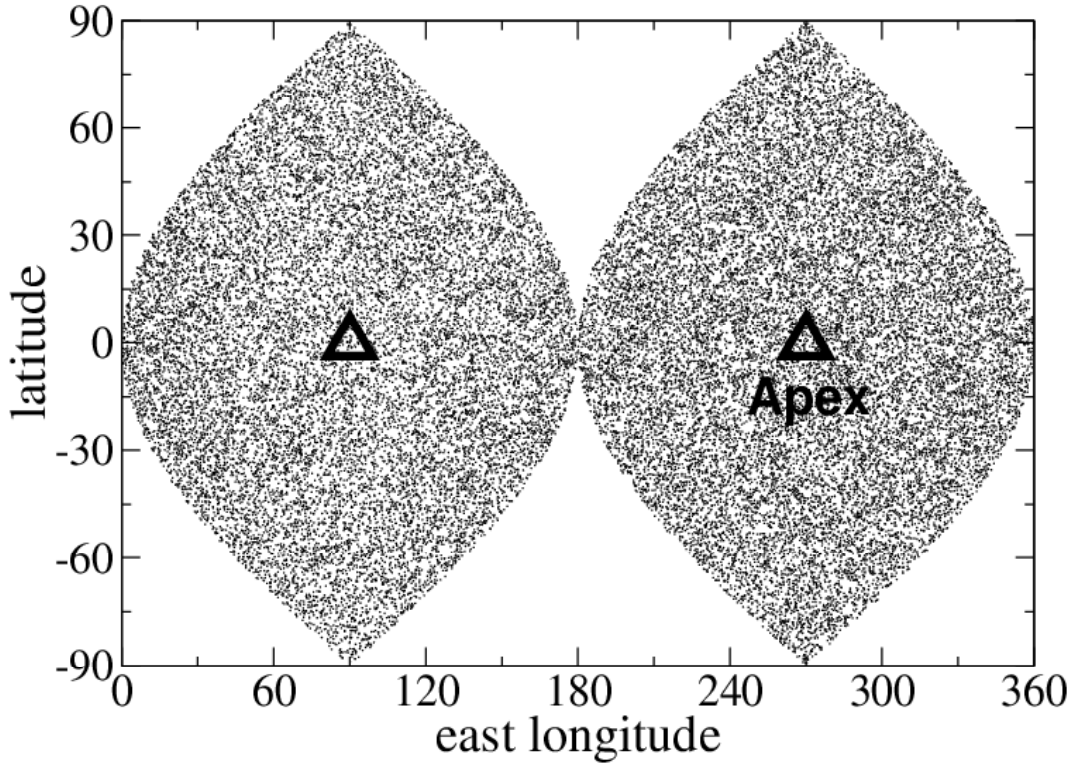,height=10.5cm,angle=0}
\end{center}
\caption{An equal-area projection of the lunar impacts from our
  simulations.  The Moon was at its current orbital distance of
  $60R_\oplus$ with 0\degrees inclination.  At the apex, one can see
  the slight enhancement in the impact density.  For clarity, only 7\%
  of the total number of impacts are shown. } 
\label{fig:hemmap}
\end{figure}

To examine crater asymmetries on the Moon, we need to convert our
simulated impacts (a sample of which are shown in
Fig.~\ref{fig:hemmap}) into craters which will account for the
added asymmetry resulting from the impact velocity $v_{imp}$ of the
impactors.  In typical crater counting studies, there is some minimum
diameter $T$ which observers are able to count down to due to image
resolution limitations.  There will be more craters larger than $T$ on
the leading hemisphere than on the trailing because leading-side
impactors have higher impact speeds on average and the
commonly-accepted scaling relation for crater size $D_c$ depends on
the velocity. 
\beq
D_c \propto v_{imp}^{2q}\;D_i^{3q}\;,
\eeq
where $q\approx 0.28-0.33$ \citep{mel89} and $D_i$ is the diameter of
the impactor.  Since both hemispheres receive flux from the same
differential impactor size distribution obeying 
\beq
\frac{dN}{dD_i} \propto D_i^{-p}\;,
\eeq
we can integrate ``down the size distribution'' to determine a weighting
factor which transforms our impacts into crater counts.  The number of
craters $N$ with $D_c > T$ produced by the impacting size distribution is  
\beq
N(D_c > T)\: =\: \int_{D_{imin}}^{\infty} \frac{dN}{dD_i}\:dD_i\:
= \: \frac{1}{p-1}\;D_{imin}^{1-p}\: \propto \: v_{imp}^{2(p-1)/3}\;,
\eeq
where $D_{imin}$ is some minimum impactor diameter and we have
substituted $D_{imin} \propto (T\;v_i^{-2q})^{1/3q}$.  Since the
differential size index $p \sim 2.8$ \citep{stu01,bottke02}, $N
\propto v_{imp}^{1.2}$.  

For each simulated impact at a specific latitude and longitude, we assign
that impact $N = C\;v_{imp}^{1.2}$ craters, where $C$
is an arbitrary proportionality constant.  The dependance on $p$ is
small as our results are essentially the same when using an
older determined value of $p = 2$.  The weak dependance arises from
the low orbital speed of the Moon relative to the encounter speeds of
the incoming projectiles.  Thus for moons such as the Galilean
satellites, whose orbital speeds are much higher compared to that of
the incoming flux, the value for $p$ becomes more important.  Note that
$D_{min}$ and $q$ are actually irrelevant to our analysis since we are
interested in only the crater numbers relative to an average rather
than the crater sizes.   

\vspace{-1.0cm}

\subsection{Latitude distribution of impacts}

\begin{figure}[H]
\begin{center}
\epsfig{file=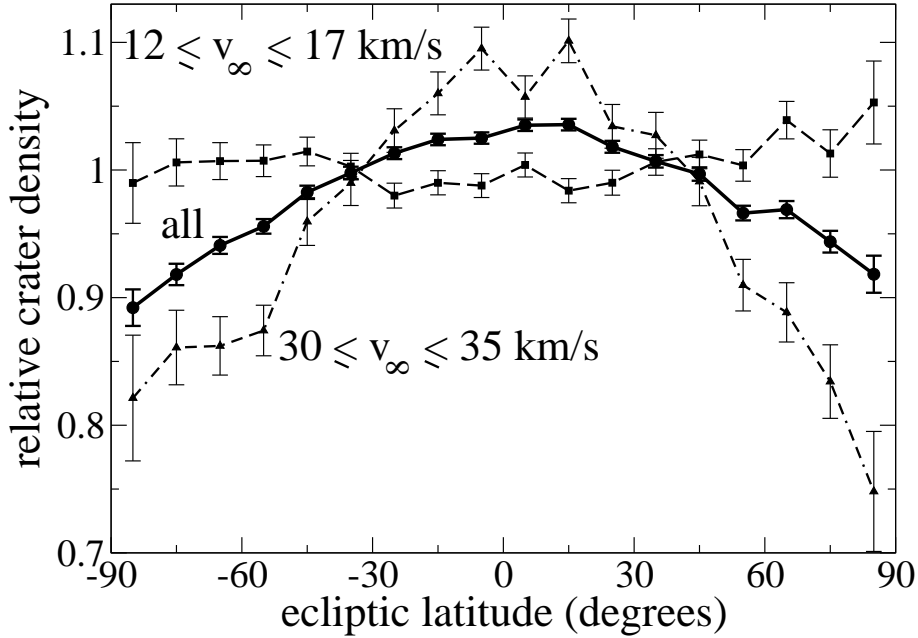,height=13.5cm,angle=-90}\vspace{-0.5cm}
\end{center}
\caption{The latitude distribution of craters on the Moon from our simulations.
  Note the scale is different here than in Fig.~\ref{fig:fevcomp1}(b)
  and that the latitudinal variation is larger for the Moon than the
  Earth.  This is because the Moon's gravity well is not deep enough
  to significantly modify incoming trajectories to higher latitudes.
  If the impacts were restricted to impactors which had $i \leq 10$\degrees,
  the variation would be larger as is the case with the terrestrial
  deliveries in Fig.~\ref{fig:fevcomp1}.  Here the Moon's inclination
  and spin obliquity are not accounted for. }  
\label{fig:moonlat}
\end{figure}

One expects the departure from uniform density in the lunar latitude
distribution to be more severe than that of the Earth due to the
Moon's smaller mass.  Comparing Fig.~\ref{fig:moonlat} to
Fig.~\ref{fig:fevcomp1}b shows this is indeed the case.  At high \vinfty
cuts, the variation in the latitude distribution tends to the predicted
cosine (see Fig.~2 in Le Feuvre and Wieczorek, 2006).  However, when
examining the real case of all lunar impacts we see only a $\sim 10$\%
($0.912 \pm 0.004$) depression at the poles when taking the ratio of
the derived crater density within 30\degrees of the poles to the
crater density in a 30\degrees band centered on the equator.  In
contrast, \citet{fev06} (their Fig.3) find a polar/equatorial ratio of
roughly 60\%.  The source of this large discrepancy is unclear since
the Moon in our simulations had zero orbital inclination and
spin obliquity, the same conditions used by \citet{fev06}, and the
latter also used the \citet{bottke02} model as an impactor source.
Despite the variation we observe being small, researchers should be
aware of this spatial variation in the crater distribution when
determining ages of surfaces (see Sec.~\ref{sec-con}).

\vspace{-1.0cm}

\subsection{Longitudinal effects}

\begin{figure}[H]
\begin{center}
\epsfig{file=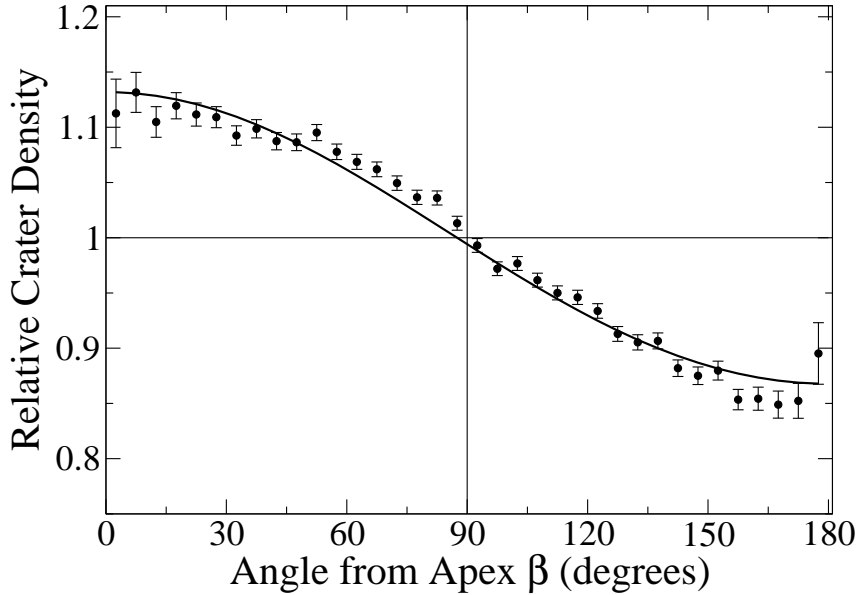,height=12.5cm,angle=-90}
\end{center}
\caption{The spatial density of simulated craters as a function of angle
  from the apex of motion, $\beta$.  The vertical axis is
  craters/$\mathrm{km}^2$ relative to the mean density over the entire
  lunar surface.  Points represent simulation results while
  the curve gives a fit using a maximum likelihood technique to the
  equation $\Gamma(\beta) = \bar{\Gamma}(1 + \alpha\cos\beta)^g$, where $g =
  2 + 1.4\;b$ and $b = 0.58$ as determined from observations.  With this
  restriction, the best fit values are $\bar{\Gamma} = 0.994$ and $\alpha =
  0.0472$ with a reduced chi-square of $\bar{\chi}_r^2 = 2.4$.}     
\label{fig:moonrad}
\end{figure}

We wish to compare the results of our numerical simulations with
observational data to create a consistent picture of the current level
of cratering asymmetry between the Moon's leading and trailing
hemispheres.  As a measure of this asymmetry, we look at the crater
density as a function of the angle away from the apex, $\beta$, which
should roughly follow the functional form of Eq.~\ref{eq:ass}.  In
Fig.~\ref{fig:moonrad} we show the results from our simulations as
well as a fit to Eq.~\ref{eq:ass} using a maximum likelihood
technique assuming a Poisson probability distribution.  The best
fit parameters resulting from an unrestricted analysis are $\bar{\Gamma} = 
1.02$, $\alpha = 0.564$, and $g = 0.225$; this would require $b =
-1.268$.  For an impactor diameter distribution following $\frac{dN}{dD}
\propto D^{-b}$, this value for the slope yields the unphysical
situation of having more large impactors than
small ones.  Obviously this cannot be correct and is a consequence of
the degeneracy present in Eq.~\ref{eq:ass}.  We once again conclude
that by fitting Eq.~\ref{eq:ass} to an observed surface distribution,
it is virtually impossible to decouple $\alpha$ and $g$ to obtain
information about the impactor size distribution and the average
encounter velocity of the impactors.  

\vspace{-1.0cm}

By counting our simulated craters, we find a value for
the GMAACA of $1.29 \pm 0.01$, significantly lower than the values of
1.4--1.7 estimated in Sec.~\ref{sec:llhe}.  This discrepancy can be 
reconciled by recalling that we find the average impactor \vinfty
to be $\sim$ 20 km/s.  Using this value in Eq.~\ref{eq:alf} and the
observationally-determined value of $b=0.58$, gives a GMAACA value of
1.32, close to the value we obtain.  In Fig.~\ref{fig:moonrad} we
notice that the distribution is rather flat for $0^\circ \leq \beta
\leq 45^\circ$ and has higher density than the predicted curve (using
$b = 0.58$) for $60^\circ \leq \beta \leq 120^\circ$.  A
$\bar{\chi}_r^2$ value of 2.4 results when the quality of fit is
assessed.  We believe the origin of this highly-significant departure
from the theoretically predicted form lies simply in the fact that
both assumptions of an (1) isotropic orbital distribution of impactors
with (2) a single \vinfty value, are violated (see
Fig.~\ref{fig:dirs}).  Thus, an observed crater field will not follow
the form of Eq.~\ref{eq:ass} in detail.      

\begin{figure}[H]
\begin{center}
\epsfig{file=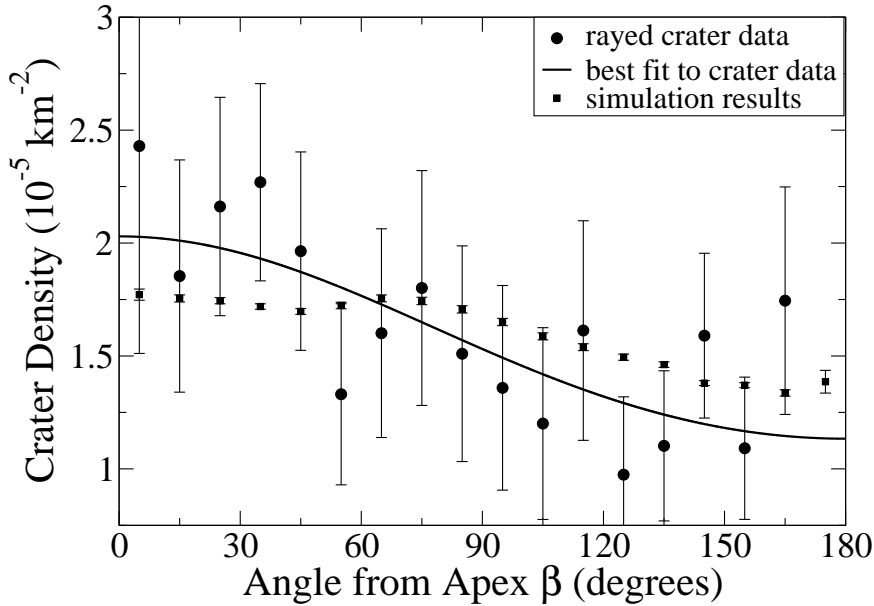,height=12.5cm,angle=-90}
\end{center} 
\caption{Density of rayed craters used in \citet{morota03}
  compared with our simulation results for the same
  lunar area and scaled to the same count (222).  The
  curve is the best fit to the rayed crater data using
  Eq.~\ref{eq:ass} as a model despite our knowledge that it is based
  on violated assumptions.  Best fit parameters are $\bar{\Gamma} =
  1.53 \times 10^-5\;\mathrm{km}^{-2},\; \alpha = 0.063,\;
  \mathrm{and}\; g = 4.62$.  With respect to the observations, the
  prediction from our numerical calculations has $\bar{\chi}_r^2 =
  0.67$.}     
\label{fig:morvsme} 
\end{figure}

Since these assumptions break down for the real impactor population,
we will instead directly compare with the rayed crater observations to
determine if the available data rule out our model.  To do this, we
scaled our craters down to 222, the same number as in the rayed crater
sample used in \citet{morota03}.  We restricted our simulated craters
to the same lunar area examined in that study.  Mare regions were
ignored as it introduces bias in rayed crater observations because
they are easier to identify against a darker background surface.  In
addition, rayed craters on mare surfaces are likely older than their
highland counterparts because it takes longer for micrometeorite
bombardment to eliminate the contrast.  Since we are interested in the
current lunar bombardment, it is necessary to eliminate this older
population.  The area sampled includes latitudes $\pm41.5$\degrees and
longitudes $70.5$\degrees $-\: 289.5$\degrees (see Fig.~1 of Morota
and Furumoto, 2003).  Using a modified chi-square test with the rayed
crater observational counts $O_i$ and the expected counts $E_i$  from
our simulations,  
\beq
\chi^2 = \sum_i^n\frac{(O_i - E_i)^2}{\sigma_{obs}^2},
\mathrm{\;\;with\;\;} \sigma_{obs} = \sqrt{O_i}   \nonumber 
\eeq
since we are dealing with Poisson statistics.  This procedure results
in a reduced chi-square value of $\bar{\chi}_r^2 = 0.67$.  Thus our
model is in excellent agreement with the observational data
(Fig.~\ref{fig:morvsme}).  

\vspace{-1.0cm}

Ideally, we would like to obtain a GMAACA value for the
rayed crater data.  However, due to small number statistics and the
restricted area (because of Mare Marginis and Mare Smythii, much of the
area near the the antapex is excluded), the  GMAACA value is poorly
measured by the available data.  Regardless, integrating the best fit 
of the rayed crater data (best fit values: $\bar{\Gamma} = 1.53 \times
10^{-5}\;\mathrm{km}^{-2},\; \alpha = 0.063,\; g = 4.62,\; \mathrm{with}\;
\bar{\chi}_r^2 = 0.49$) from 0-30\degrees and 150-180\degrees to form
the GMAACA ratio gives $1.7_{-0.6}^{+0.9}$, consistent with our result.

\vspace{-0.5cm}
\subsection{Varying the Earth-Moon distance and the effect of inclination}
\begin{figure}[H]
\begin{center}
\epsfig{file=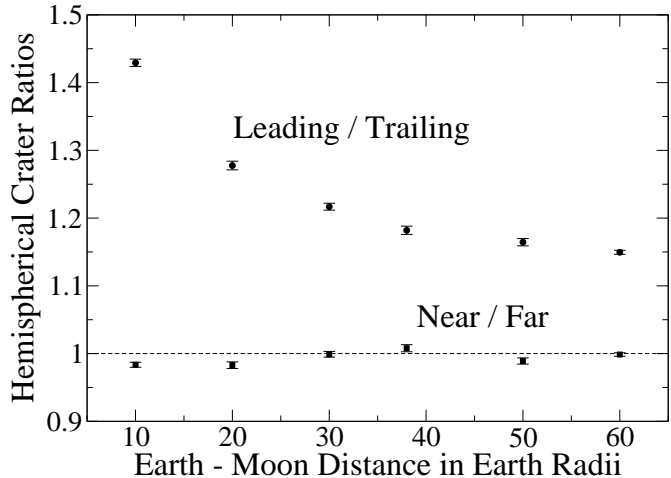,height=10.0cm,angle=-90}
\end{center} 
\caption{The ratio between number of craters on the leading
  versus trailing hemispheres and the same ratio between the nearside
  and farside hemispheres as a function of lunar orbital distance.  As
  expected, smaller orbital distances increase the asymmetry between
  leading and trailing hemispheres.  This is a result of the increased
  orbital speed as the Moon is brought closer to the Earth.  For all
  lunar distances there is minimal nearside/farside asymmetry, with
  some evidence the Earth shielded the lunar nearside in the very
  distant past when the lunar semimajor axis was $<\; 25\;R_\oplus$. }
\label{fig:dist}
\end{figure}

Due to tidal evolution, in the distant past the Moon's orbit was
smaller.  As the orbital distance is decreased, the Moon's orbital
speed rises.  This increases the impact speeds on the leading
hemisphere and makes ``catching up'' to the Moon from behind more
difficult.  As Eq.~\ref{eq:alf} suggests, the degree of apex/antapex
asymmetry on the Moon is expected to increase as the orbital speed of
the satellite does.  We examined this by running other simulations
with an Earth-Moon separation of $a = $50, 38, 30, 20, and 10
$R_\oplus$.  Figure~\ref{fig:dist} shows our results.  Assuming the same
impactor orbital distribution in the past, only a mild increase in
the apex enhancement is seen (since the orbital speed only increases
as $1/\sqrt(a)$).  The increase is that expected based on the
resulting change in $\alpha$ caused by the larger $v_{orb}$ (see
Eq.~\ref{eq:alf}).     

\vspace{-1.0cm}

As discussed in Sec.~\ref{sec-nearfar}, the asymmetry
between the near and far hemispheres should depend on the lunar 
distance.  Figure~\ref{fig:dist} shows the ratio between nearside and
farside craters from our simulations.  For all lunar distances we find
very little asymmetry.  Thus, our results do not support the most
recent study which claims a factor of four enhancement on the nearside
when compared to the far \citep{fev05}, but are in good agreement with
the work done by \citet{wiesel71} and \citet{band73}.  We see mild
evidence for Bandermann and Singer's (1973) assertion of the Earth
acting as a shield for $a < 25 R_\oplus$ and little effect outside
this distance. Thus, Bandermann and Singer's estimate of no measurable
nearside/farside asymmetry in the current cratering rate is correct. 

In the bulk of our simulations we have used the approximation
that the Moon's orbit is in the ecliptic plane.  Since we show that
the latitudinal dependence of lunar cratering is weak ($\sim$ 10\%
reduction within 30$^{\circ}$ of the poles relative to a 30\degrees
band centered on the Moon's equator), we do not expect the inclusion of
the moon's orbital inclination to alter our results significantly,
although we expect the polar asymmetry to monotonically decrease with
increasing orbital inclination. We have confirmed this by computing
the GMAACA and polar asymmetry ratios for a less-extensive set of
simulations in which the lunar orbital inclination is initially set to
its current value of 5.15$^{\circ}$ and we use the sub-Earth point at
the time of impact to compute lunocentric latitudes and longitudes.
We find a slight reduction of GMAACA to $1.24\pm0.02$ from
($1.29\pm0.01$) and a crater density within 30$^{\circ}$ of the pole
that is statistically the same as the 0\degrees inclination case
($0.914 \pm 0.009$ instead of $0.912 \pm 0.004$).

\section{Summary and conclusions}
\label{sec-con}

We have used the debiased NEO model of \citet{bottke02} to examine the
bombardment of the Earth-Moon system in terms of various impact and
crater asymmetries.  For Earth arrivals we find a $< 1$\% variation in
the ratio between the areal densities within 30\degrees of the poles and
within a 30\degrees band centered on the equator.  The local time
distribution of terrestrial impacts from NEOs is enhanced during the
AM hours.  While this fall-time distribution corresponds well to
recent radar data, it is in disagreement with the chondritic meteorite
data and their derived pre-atmospheric orbital distributions.  This
discrepancy thus reinforces the conclusion of \citet{morglad98} that the
large amount of decimeter-scale material being ejected from the main
asteroid belt onto Earth-crossing orbits must be collisionally
depleted before much of it can evolve to orbits with $a < 1.5$~AU.

A significant result is that we find the average impact speed
onto the Moon to be $\bar{v}_{imp} = 20$~km/s, with a non-negligible
higher-speed tail
(Fig.~\ref{fig:velhist}).
This combined with quantification of the non-uniform surface cratering
has implications for both tracing crater fields back to the size
distribution of the impactors and the absolute (or relative) dating of
cratered surfaces. First, the higher impact speeds we find mean that
lunar impact craters (at least in the post-mare era when we believe
the NEO orbital distribution we are using is valid) have been produced
by smaller impactors than previously calculated.  This roughly 10\%
higher average impact speed corresponds to lunar impactors which are
10\% smaller on average than previously estimated; this small
correction has ramifications for proposed matches between lunar crater
size distributions and impactor populations (\eg Strom \etal 2005). 
%ANOTHER?

We find two different spatial asymmetries in current crater production
due to NEOs.  As expected, due to its smaller mass, the Moon exhibits more
latitudinal variation ($\sim 10$\%) in our simulations than the Earth.
When comparing our simulation results to young rayed craters on the
Moon, the surface density variation we predict is completely consistent
with available data; we obtain a leading versus trailing asymmetry
of $1.29 \pm 0.01$ (GMAACA value), which corresponds to a 13\%
increase (decrease) in crater density at the apex (antapex) relative
to the average.  These results indicate that using a single
globally-averaged lunar crater production could give ages in error by
up to 10\% depending on the location of the studied region.  For
example, post-mare studies of the Mare Orientale region would
overestimate its age by $\sim10$\% due to its proximity to the apex,
assuming that leading point and poles of the Moon have not changed
over the last $\sim$4~Gyr.  Similarly, the degree of bombardment on
Mare Crisium (not far from the antapex) would be lower than the global
average.  These effects will only be testable for crater fields with
hundreds of counted craters so that the Poisson errors are small
compared to the 10\% variations we find; in most studies of
``young'' ($<4$ Gyr) lunar surfaces the crater statistics are poorer than
this (\eg St\"offler and Ryder 2001). 

When the orbital distance of the Moon was decreased (as it
was in the past because of its tidal evolution), the ratio between
simulated craters on the leading hemisphere versus the trailing  
increased as expected due to the higher orbital speed of the
satellite at lower semi-major axes.
We find virtually no nearside/farside asymmetry
until the Earth-Moon separation is less than 30 Earth-radii,
which under currently-accepted lunar orbital evolution models dates
to the time before 4~Gyr ago (at which point the current NEO
orbital distribution may not be a good model for the impactors).
Interior to 30 Earth-radii we find that the Earth serves as a
mild shield, reducing nearside crater production by a few percent.

\bigskip
\centerline{\bf Acknowledgments}
\medskip

Computations were performed on the LeVerrier Beowulf cluster at UBC,
funded by the Canadian Foundation for Innovation, The BC Knowledge
Development fund, and the UBC Blusson fund.  BG and JG thank NSERC for
research support.  

\singlespace
%--------------------------------------------------
% Tables

\newpage
\newpage
%--------------------------------------------------

%-----------------------------------------------------------------
% Figures and captions:

\end{document}